\documentclass{emulateapj}
\usepackage{epsf,graphics}
\usepackage{subfigure,graphicx}
\usepackage{bm}

\begin{document}

\title{Binary populations in Milky Way satellite galaxies: constraints from 
multi-epoch data in the Carina, Fornax, Sculptor, and Sextans dwarf spheroidal 
galaxies}

\def\mail{*}

\author{Quinn E. Minor}
\affil{Department of Science, Borough of Manhattan Community College, City 
University of New York, New York, NY 10007, USA}
\affil{Department of Astrophysics, American Museum of Natural History, New York, NY 10024, USA}
\email{qminor@bmcc.cuny.edu}

\submitted{Submitted to ApJ 2013-05-25; accepted 2013-11-06}

\begin{abstract}
We introduce a likelihood analysis of multi-epoch stellar line-of-sight 
velocities to constrain the binary fractions and binary period distributions of 
dwarf spheroidal galaxies.  This method is applied to multi-epoch data from the 
Magellan/MMFS survey of the Carina, Fornax, Sculptor and Sextans dSph galaxies, 
after applying a model for the measurement errors that accounts for binary 
orbital motion. We find that the Fornax, Sculptor, and Sextans dSphs are 
consistent with having binary populations similar to that of Milky Way field 
binaries to within 68\% confidence limits, whereas the Carina dSph is 
remarkably deficient in binaries with periods less than $\sim$10 years. If Carina 
is assumed to have a period distribution identical to that of the Milky Way 
field, its best-fit binary fraction is $0.14^{+0.28}_{-0.05}$, and is 
constrained to be less than 0.5 at the 90\% confidence level; thus it is 
unlikely to host a binary population identical to that of the Milky Way field.  
By contrast, the best-fit binary fraction of the combined sample of all four 
galaxies is $0.46^{+0.13}_{-0.09}$, consistent with that of Milky Way field 
binaries. More generally, we infer probability distributions in binary 
fraction, mean orbital period, and dispersion of periods for each galaxy in the 
sample. Looking ahead to future surveys, we show that the allowed parameter 
space of binary fraction and period distribution parameters in dSphs will be 
narrowed significantly by a large multi-epoch survey. However, there is a 
degeneracy between the parameters that is unlikely to be broken unless the 
measurement error is of order $\sim$0.1 km~s$^{-1}$ or smaller, presently 
attainable only by a high-resolution spectrograph.
\end{abstract}


\section{Introduction}\label{sec:intro}

Recent progress in hydrodynamical simulations has allowed, for the first time, 
numerical simulations of star formation that include all the relevant physics 
down to solar system scales (\citealt{bate2009}; \citealt{offner2009}). Star 
formation is now understood to occur due to the gravitational collapse and 
fragmentation of a turbulent molecular cloud, generally in the presence of 
radiative feedback (\citealt{bate2012}) and magnetic fields 
(\citealt{price2010}). Simulations of star formation that incorporate these 
physical effects now offer detailed predictions about the statistical 
properties of stellar systems that must be tested against observational data to 
arrive at a complete theory of star formation. A major component of such 
predictions involve the statistical properties of binary star systems.  While 
approximately half of solar-type field stars in the Milky Way are known to have 
binary companions, the fraction of young pre-main sequence stars in binary or 
higher-order star systems is much higher and possibly greater than 90\% 
(\citealt{leinert1993}; \citealt{kohler1998}; \citealt{patience2002}). This 
binary fraction is later reduced as many wide binaries are subsequently 
disrupted by cluster dynamics or orbital decay. Thus, binary systems are likely 
the dominant mode of star formation, and the ability to successfully predict 
their statistical properties will be an essential component of a complete 
theory of star formation.

At present, a complete census of the statistical properties of binary star 
systems exists only in the solar neighborhood, at distances out to $\approx$ 30 
pc from the Sun (\citealt{raghavan2010}; \citealt{duquennoy1991}). While a 
number of visual binary studies have been made in nearby open clusters 
(\citealt{brandner1998}; \citealt{patience2002}), these studies are limited to 
wide binaries with separations $\gtrsim$ 10 AU. Moreover, such wide binaries 
are likely to have been ``dynamically processed'' by stellar encounters within 
the cluster, and thus may not represent the primordial binary populations 
present shortly after star formation has occurred. Likewise, spectroscopic and 
photometric studies have been made in globular clusters (\citealt{yan1996}; 
\citealt{milone2012}; \citealt{sollima2007}; \citealt{sollima2012}), but in 
many of these systems even close binaries have been dynamically processed in 
the dense core of the cluster over cosmological timescales.  To test theories 
of star formation, it is important to study binary populations in diffuse 
clusters or field populations over a range of separations for which the binary 
systems can be considered primordial, and compare their statistical properties 
to that of simulations.

Dwarf spheroidal (dSph) galaxies are excellent objects for this purpose.  
Because they are more diffuse than large globular clusters, stellar encounters 
are sufficiently rare in dSphs that binary stars with separations less than 10 
AU should be largely unaltered since their formation. Dwarf spheroidals 
represent a variety of star forming environments, ages, and metallicities, 
against which theories of star formation can be tested.  Furthermore, as 
simulations advance and our understanding of star formation is refined, dSphs 
may ultimately become laboratories for a new brand of ``stellar archaeology'': 
using the statistical properties of their binary star populations to discern 
the initial conditions under which the galaxy originally formed. The extent to 
which binary properties are sensitive to initial conditions during star 
formation is still an open question.  While differences between binary 
populations of different clusters have been observed (\citealt{king2012}), 
these differences may be largely due to subsequent dynamical processing with 
the cluster (\citealt{marks2012}, \citealt{kroupa1995}).  If binary star 
properties do indeed have a significant dependence on initial conditions (e.g. 
cloud temperature, density, magnetic field, metallicity), they may eventually 
become a tool for understanding star formation histories.

Another reason for constraining binary properties in dwarf galaxies is to 
understand the effect of binary orbital motion on the stellar velocity 
distribution, and hence the estimated mass distribution, of the galaxies 
(\citealt{olszewski1996}; \citealt{minor2010}).  The spatial mass distribution 
of dark matter within a galaxy, including the existence and size of a central 
core, is determined by properties of the dark matter particle itself.  However, 
because there exists a degeneracy between mass and velocity anisotropy 
(\citealt{wolf2010}), it is not possible to infer the dark matter density 
profile in a straightforward way. One way to overcome this degeneracy is to 
make use of higher moments in the velocity distribution to distinguish between 
radial and tangential orbits, and hence infer the velocity anisotropy 
(\citealt{lokas2005}; \citealt{lokas2009}; \citealt{richardson2013}).  However 
because binary motion can also contribute significantly to higher moments in 
the velocity distribution, it is important to correct for binaries when 
determining the velocity anisotropy. Therefore, quantifying the effect of 
binary motion on a galaxy's velocity distribution can lead to a better 
understanding of the distribution of dark matter in dwarf galaxies.

Finding detailed constraints on the binary populations in these galaxies is a 
difficult task.  In the most sensitive spectroscopic surveys of dwarf 
spheroidals to date, typical line-of-sight velocity measurement errors 
($\approx$ 1-3 km~s$^{-1}$ for red giant stars) are large enough to render binary 
motion in systems with periods longer than 10 years unobservable.  Since 
binaries with such long periods cannot be directly observed in these galaxies, 
any statistical information about long-period binaries can only be inferred by 
extrapolation.  Furthermore, among the red giant population, information about 
the shortest-period binaries is lost because close binary systems are destroyed 
when the red giant star overflows its Roche lobe (\citealt{paczynski1971}).  
Another complication is that any component of a star's velocity due to binary 
motion must be separated from the measurement error, and therefore deriving 
accurate and reliable measurement errors is essential before obtaining binary 
constraints.

In this paper, we show that all of these difficulties can be surmounted by a 
likelihood approach, and demonstrate the method using multi-epoch data from the 
Magellan/MMFS sample of \cite{walker2009} in the Carina, Fornax, Sculptor, and 
Sextans dwarf spheroidal galaxies. In Section \ref{sec:likelihood} we derive a 
multi-epoch likelihood for binary and non-binary stars, followed by a 
discussion of binary model uncertainties in Section 
\ref{sec:binary_properties}.  In Section \ref{sec:error_model}, we show how 
this likelihood can incorporate an error model to derive accurate measurement 
errors in the Magellan/MMFS sample, which we apply to the data in Section
\ref{sec:error_model_data}. To ensure the error model has properly reproduced 
the measurement errors, we then test the model on simulated data in Section 
\ref{sec:test_error_model}.  Next, the binary fraction in each galaxy is 
inferred in Section \ref{sec:binary_fraction}, under the assumption of a Milky 
Way-like period distribution, and our main results are plotted in Figure 
\ref{fig:bfposts_alldwarfs_sn1.2}, wherein probability distributions in binary 
fraction are plotted for each galaxy. The inferred binary fraction of the 
combined sample of all four galaxies is plotted in Figure 
\ref{fig:bfposts_combined_dsphs}, and the best-fit binary fractions in each 
galaxy are listed in Table \ref{tab:binary_fraction}. In Section 
\ref{sec:period_distribution} we find more general constraints on the binary 
fraction, mean period, and spread of periods in each galaxy.  In Section 
\ref{sec:simulations} we apply our methodology to simulated data to investigate 
the prospects for finding more precise binary constraints in a larger 
multi-epoch dataset. Finally, we discuss implications of our results in Section
\ref{sec:discussion} and conclude in Section \ref{sec:conclusion}.

\section{Multi-epoch likelihood}\label{sec:likelihood}

In order to constrain binary properties from radial velocity measurements, we 
must find the likelihood of our binary model parameters for a given set of 
measured velocities.  If the fraction of stars in 
binary systems $B$ is taken as a model parameter, then the likelihood will take the form $\mathcal{L} = 
\mathcal{L}_{nb} + B\mathcal{L}_b$, where $\mathcal{L}_{nb}$ and 
$\mathcal{L}_b$ represent the likelihood for non-binary and binary stars, 
respectively. In addition, we will choose a set of 
parameters $\mathcal{P}$ that specify the distributions of binary 
properties. These binary properties may include the orbital periods, mass 
ratios and orbital eccentricities. The assumed distributions in these properties and our 
chosen model parameters will be described in Section 
$\ref{sec:binary_properties}$.

In principle, even in a sample of stars with only one velocity measurement 
each, binary properties could be inferred from the line-of-sight velocity 
distribution of the sample, where the presence of binary stars is evident from a
non-Gaussian tail representing short-period binaries with high orbital 
velocities. However, inferring binary properties from the velocity distribution 
is fraught with difficulties.  Because the classical dwarf spheroidal galaxies 
have velocity dispersions greater than 7 km~s$^{-1}$, only stars with radial 
velocities that differ from the galaxy's systemic velocity by at least 
$\gtrsim$ 14 km~s$^{-1}$ will be evident as binary stars, and only a small 
fraction of binaries have velocity amplitudes this large.  For this reason, 
extremely large samples would be required to have any hope of constraining 
binary properties from single-epoch velocity measurements alone.  Contamination 
from Milky Way stars must also be accounted for, as such stars may be mistaken 
for binaries.  Finally, non-Gaussianities may be present in the galaxy's 
velocity distribution due to velocity anisotropy and must also be distinguished 
from binary motion.

Because the present work is primarily concerned with finding binary 
constraints, we can avoid the difficulties inherent in using the galaxy's 
velocity distribution by focusing on the stars with repeat measurements.  
Suppose a star has $n$ radial velocity measurements 
$\{v_i\}=\{v_1,\cdots,v_n\}$ and measurement errors $\{\sigma_i\}$ taken at the 
corresponding times, $\{t_i\}$.
For stars with at least two measurements ($n \geq 2$), we shall derive a 
likelihood in velocity \emph{differences} $\Delta v_1$, $\Delta v_2$, and so 
on, where we define $\Delta v_i = v_{i+1} - v_1$.  For the sake of readability, we will henceforth suppress the brackets denoting sets 
of measurements (e.g., $P(\{v_i\}) \rightarrow P(v_i)$).

Our desired likelihood is derived as follows. We first write down the 
probability, for single stars, of obtaining a set of radial velocity measurements 
$\{v_i\}$ assuming that it has intrinsic velocity $v_{true}$. Assuming the 
measurement errors are Gaussian, this likelihood can be expressed as follows:

\begin{eqnarray}
\label{eq:v_given_vcm_like}
P(v_i|v_{true},\sigma_i) & = & \prod_{i=1}^n 
\frac{e^{-(v_i-v_{true})^2/2\sigma_i^2}}{\sqrt{2\pi\sigma_i^2}} \nonumber \\
& = & \mathcal{N}(v_i,\sigma_i) \frac{e^{-(v_{true} - \langle v 
\rangle)^2/2\sigma_m^2}}{\sqrt{2\pi\sigma_m^2}} \nonumber \\
\end{eqnarray}
Here, $\langle v\rangle$ and $\sigma_m$ are the variance-weighted average 
velocity and measurement error,
\begin{equation}
\langle v \rangle = \sigma_m^2 \sum_{i=1}^n \frac{v_i}{\sigma_i^2},
\label{eq:average_v}
\end{equation}

\begin{equation}
\sigma_m^2 = \left(\sum_{i=1}^{n}\frac{1}{\sigma_i^2}\right)^{-1},
\label{eq:sigm}
\end{equation}
and a tedious algebraic calculation shows that the factor 
$\mathcal{N}(v_i,\sigma_i)$ is given by

\begin{eqnarray}
\lefteqn{
\mathcal{N}(v_i,\sigma_i) = \frac{\sqrt{2\pi\sigma_m^2}}{\prod_{i=1}^n \sqrt{2\pi\sigma_i^2}} } \nonumber\\
& \times & \exp\left\{\frac{-1}{4}\sum_{i,j=1}^n\frac{(v_i-v_j)^2}{\sigma_i^2 + \sigma_j^2 + \sigma_i^2 \sigma_j^2\left(\sum_{k \neq i,j}\frac{1}{\sigma_k^2}\right) } \right\}.
\end{eqnarray}
The final term in the denominator of the exponent is explicitly zero when $n=2$.  
Note that this factor depends only on velocity differences $v_i-v_j$; it has no 
dependence on the star's intrinsic velocity $v_{true}$.

To derive a likelihood in velocity differences $\Delta v$, we change variables 
in Equation \ref{eq:v_given_vcm_like} from ($v_1$,$v_2$,$v_3$, \ldots) to ($\langle 
v \rangle$,$\Delta v_1$,$\Delta v_2$, \ldots) and integrate over $\langle v 
\rangle$. Then the Gaussian factor in Equation \ref{eq:v_given_vcm_like} integrates 
to 1, and we are left with

\begin{eqnarray}
\lefteqn{
\mathcal{L}_{nb}(\Delta v_i|\sigma_i) = \frac{\sqrt{2\pi\sigma_m^2}}{\prod_{i=1}^n \sqrt{2\pi\sigma_i^2}} } \nonumber\\
& \times & \exp\left\{\frac{-1}{4}\sum_{i,j=1}^n\frac{(\Delta v_{i-1}-\Delta v_{j-1})^2}{\sigma_i^2 + \sigma_j^2 + \sigma_i^2 \sigma_j^2\left(\sum_{k\neq i,j}\frac{1}{\sigma_k^2}\right) } \right\}
\label{eq:nonbinary_likelihood}
\end{eqnarray}

This is our desired likelihood for non-binary stars; it is 
essentially the same factor $\mathcal{N}(v_i,\sigma_i)$ above, but expressed in 
the new variables $\Delta v_i$. As an aside, we note that it is possible to 
derive a likelihood in \emph{all} the velocities $v_i$ using this formalism; 
this is the approach taken in \cite{martinez2011}.  Since we are only 
interested in binary constraints rather than the galaxy's intrinsic velocity 
distribution, only a likelihood in velocity differences is considered in the 
present work.

The corresponding likelihood for binary stars cannot be found analytically, so 
it must be calculated by running a Monte Carlo simulation for a large number of 
simulated stars and binning in the velocity differences $\Delta v_i$ (the 
details of the Monte Carlo simulation are discussed in \citealt{minor2010}).  
The likelihood depends not only on the velocities and measurement errors, but 
also the time intervals between epochs and the absolute magnitude $M_V$ of the star, from 
which we determine the mass and radius of the primary star via a stellar 
population synthesis model. The likelihood will also depend on the set of 
parameters $\mathcal{P}$ used to characterize the binary population which we 
discuss further in Section \ref{sec:binary_properties}. In terms of all the 
relevant model parameters, the likelihood for each star is therefore

\begin{eqnarray}
\lefteqn{
\mathcal{L}(\Delta v_i|\sigma_i,t_i,M;B,\mathcal{P}) } \nonumber\\
& = & (1-B)\mathcal{L}_{nb}(\Delta v_i|\sigma_i) + B\mathcal{L}_b(\Delta 
v_i|\sigma_i,t_i,M;\mathcal{P}).
\label{eq:full_binary_likelihood}
\end{eqnarray}

We can put this in a somewhat simpler form by noting that the likelihood for 
single stars does not depend on any model parameters, and thus can be divided 
out of Equation \ref{eq:full_binary_likelihood}. We then have

\begin{equation}
\mathcal{L}(\Delta v_i|\sigma_i,t_i,M;B,\mathcal{P}) \propto (1-B) + B J(\mathcal{P}),
\label{eq:jfactor_eqn}
\end{equation}
where
\begin{equation}
J(\mathcal{P}) = \frac{\mathcal{L}_b(\Delta v_i|\sigma_i,t_i,M;\mathcal{P})}{\mathcal{L}_{nb}(\Delta v_i|\sigma_i)}
\end{equation}

For non-binary stars or binaries with small velocity amplitudes compared to the 
measurement errors, the $J$-factor in Equation \ref{eq:jfactor_eqn} will be of order 
1 or less, since the binary likelihood will typically be slightly smaller than 
the non-binary likelihood.  However, if a binary star exhibits large velocity 
changes compared to the measurement errors, the $J$-factor will be larger than 
1, possibly by many orders of magnitude, since the non-binary likelihood 
$\mathcal{L}_{nb}$ in the denominator will be very small.

For each star, the factor $J(\mathcal{P})$ can be calculated over a grid of 
values in the binary parameters $\mathcal{P}$, after which the $J$-factors can 
then be found by interpolation for any values of the binary parameters 
$\mathcal{P}$. After calculating the $J$-factors, the binary fraction $B$ and 
parameters $\mathcal{P}$ can then be inferred either by a maximum-likelihood or 
Bayesian analysis.

\section{Binary population model}\label{sec:binary_properties}

Here we confront the issue of which binary population model parameters to constrain and how 
to deal with uncertainties in these parameters.  In addition to the binary fraction, 
binary star populations are described by distributions in three 
properties: the orbital period $P$, mass ratio $q$, eccentricity $e$. (We 
have assumed a uniform distribution of orbital inclinations, as is observed in 
Milky Way binaries; see \citealt{minor2010}).  Without a large set of 
measurements, orbital eccentricities are difficult to constrain because binaries with 
high eccentricity spend only a small fraction of their orbital period 
near periastron where their observed velocities are high. For this reason, we 
will assume the distribution of eccentricities to have the form given in 
\cite{minor2010}, with parameters fixed to those observed in solar 
neighborhood binaries.

Similarly, several measurements are typically required to constrain the 
mass ratio of a binary separately from its orbital period.  On the other hand, 
theoretical considerations and open cluster surveys suggest that while the 
period distributions of binary populations may vary significantly depending 
on initial conditions (\citealt{fisher2004}, \citealt{brandner1998}), the 
distribution of mass ratios may be approximately universal in form.  This is 
known to be the case for binaries with long periods ($P > 1000$ days), where the distribution of mass ratios
traces the Salpeter initial mass function to good approximation for $q \gtrsim 
0.5$ (assuming the primary star masses fall within a restricted range, 
as it does for stellar populations older than a few Gyr; cf.  
\citealt{duquennoy1991}).  The mass ratio distribution observed in 
shorter-period binaries is closer to being uniform 
(\citealt{goldberg2003}; \citealt{mazeh1992}), although it is uncertain whether 
this distribution is universal in form. In light of the above considerations, 
we assume the distribution of mass ratios to take a fixed form similar 
to that observed in the solar neighborhood, described in \citet{minor2010}, 
with a Gaussian form for long-period binaries and a flat distribution for 
short-period binaries.

An important caveat is that if the mass distribution differs from our assumed 
form, this can introduce bias in the results. For example, if the actual mass 
distribution for short-period binaries rises steeply at high mass ratios, 
instead of being flat, this would result in greater binary velocity variations.  
Therefore our analysis would infer a somewhat higher binary fraction than is 
actually present under the assumption of a flat mass ratio distribution. In 
principle, we could also vary the parameters describing the mass ratio 
distribution to account for this, but this is computationally intensive; the 
binary likelihoods will have to be computed over a grid of binary parameters, 
including the parameters used to describe the mass ratio distribution. In 
addition, without a large number of epochs to separate each star's binary 
period from its velocity amplitude, these parameters will be highly degenerate 
with other binary parameters (specifically, the binary fraction and period 
distribution parameters) and thus we are unlikely to infer anything about the 
behavior of the mass-ratio distribution. This degeneracy is similar to that of 
the period distribution parameters and binary fraction, which will be discussed 
in detail in Section \ref{sec:simulations}.

By way of analogy to Milky Way field binaries, we assume the period 
distribution of dwarf spheroidal galaxies to have a log-normal form
(\citealt{duquennoy1991}; \citealt{fischer1992}; \citealt{mayor1992}; 
\citealt{raghavan2010}). Initially we shall fix the mean period $\mu_{\log P}$ 
and dispersion of periods $\sigma_{\log P}$ to the values observed in solar 
neighborhood binaries (where $\mu_{\log P}=2.24$, $\sigma_{\log P} = 2.3$), 
while in later sections (\ref{sec:period_distribution}-\ref{sec:simulations}) 
these parameters will be allowed to vary.  In general, we therefore have three 
binary parameters that will be constrained: the binary fraction $B$, the mean 
log-period $\mu_{\log P}$, and log-spread of periods $\sigma_{\log P}$. It 
should be borne in mind that for the dwarf galaxy samples considered in this 
paper, binary motion in stars with periods $\gtrsim$ 10 years will be 
unobservable due to the measurement error (the correspondence between 
observable binary periods and measurement error will be explored in detail in 
Section \ref{sec:required_error}).  Thus we can only directly probe the period 
distribution for periods shorter than $\sim$10 years, so the shape of the 
period distribution at longer periods must be inferred by extrapolation.  This 
extrapolation will hold only to the extent that the period distribution takes a 
symmetric log-normal form, which we take as a working hypothesis in this paper.

\section{Measurement error model}\label{sec:error_model}

Since any velocity variations beyond what is indicated by the measurement 
errors may be interpreted as binary motion, to constrain binary properties it 
is essential to estimate the measurement errors as well as possible. In the 
data from \cite{walker2009}, measurement errors are determined from repeat 
velocity measurements by applying an error model, which will be reviewed and 
extended in this section.  It is important to note, however, that if 
measurement errors are estimated by means other than velocity variability (e.g.  
from S/N properties), an error model such as the one considered 
here need not be applied before obtaining binary constraints.

In the error model from \cite{walker2009}, Gaussian measurement errors $\sigma$ 
are estimated using the model

\begin{equation}
\sigma = \sqrt{\sigma_0^2 + \sigma_{CCF}^2},
\label{eq:walker_error_model}
\end{equation}

\begin{equation}
\sigma_{CCF} = \frac{\alpha}{(1+R)^x}
\label{eq:walker_sigma_ccf}
\end{equation}
where $\sigma_0$ is a baseline error, $\sigma_{CCF}$ is the spectrum 
cross-correlation error, and $R_i$ is called the Tonry-Davis $R$-value for the 
measurement \citep{tonry1979}, defined as the height of the maximum peak in the 
spectrum cross-correlation function divided by the average peak height. This 
error model is an extension of the error model from \cite{tonry1979}, which 
takes $x=1$.  The MIKE spectrograph has two channels, ``red'' and ``blue'', 
which operate over different wavelength ranges, so we must have two sets of 
error model parameters representing each channel.  Our error model then has 
four parameters given by the set $\mathcal{E} = \{ \alpha_{red}$, 
$\alpha_{blue}$, $x_{red}$ and $x_{blue}\}$.  The baseline error also depends 
on the channel color, and was determined to be $\sigma_{0,red} = 0.6$ km~s$^{-1}$ and 
$\sigma_{0,blue} = 0.26$ km~s$^{-1}$. 

The implementation of the error model described above in \cite{walker2007} has 
a few disadvantages. First, their method of constraining the error model 
parameters in makes certain assumptions about the intrinsic velocity of each 
star \emph{before} deriving measurement errors.  Since the star's intrinsic 
velocity cannot be well determined from repeat measurements without knowing the 
measurement errors in the first place, this may lead to bias in the error 
model.

In contrast, since our multi-epoch likelihood is expressed solely in terms of 
velocity differences, we can naturally incorporate the error model with no 
assumptions about intrinsic velocity. If the star's measurements all have time 
intervals of less than a week between them, binary motion can be neglected and 
we can incorporate the error model into our likelihood by expressing the 
measurement error in terms of the error model parameters. Combining 
Equations \ref{eq:walker_error_model} and \ref{eq:walker_sigma_ccf}, we can write 
the error in the $i$th measurement of the star as

\begin{equation}
\sigma_i = \sqrt{\sigma_{0,c_i}^2 + \frac{\alpha_{c_i}^2}{(1+R_i)^{2x_{c_i}}}},
\end{equation}
where $c_i$ is the channel color of the $i$th measurement (either ``red'' or 
``blue''). This expression is then inserted into the likelihood from 
Equation \ref{eq:nonbinary_likelihood}, which we now write as

\begin{eqnarray}
\label{eq:nonbinary_likelihood_errmodel}
\lefteqn{
\mathcal{L}(\Delta v_i|R_i,c_i; \mathcal{E}) = \frac{\sqrt{2\pi\sigma_m^2}}{\prod_{i=1}^n \sqrt{2\pi\sigma_i^2}} } \nonumber\\
& \times & \exp\left\{\frac{-1}{4}\sum_{i,j=1}^n\frac{(\Delta v_{i-1}-\Delta 
v_{j-1})^2}{\sigma_i^2 + \sigma_j^2 + \sigma_i^2 \sigma_j^2\left(\sum_{k\neq 
i,j}\frac{1}{\sigma_k^2}\right) } \right\}.
\end{eqnarray}

More importantly, the error model in \cite{walker2007} does not take binary 
motion into account, which may bias the errors if the time between measurements 
is longer than a few weeks. We shall address this by incorporating binarity 
into the likelihood of each star if at least one of the time intervals between 
measurements is long enough for binary velocity variation to be observed. To 
determine this minimum time interval, we note that for intervals smaller than 
$\sim$ 10 days, the distribution of observed velocity changes is 
well-approximated by a Gaussian; thus we choose 10 days as our minimum 
threshold for observed binary motion.  Since we are now using the binary 
likelihood, we must constrain the binary fraction $B$ and the error model 
parameters simultaneously.

At this stage, the reader may be wondering why our binary model is being 
applied \emph{twice}: first to find the measurement errors, then afterward to 
constrain the binary fraction $B$ and binary model parameters $\mathcal{P}$.  
Why not constrain all the error model and binary parameters simultaneously? The 
answer is that this would be extremely computationally expensive, because the 
binary likelihoods must be recalculated every time the error model parameters 
are varied.  Fortunately, if we are (for the moment) only interested in 
deriving the best possible measurement errors, it is sufficient to assume a 
Milky Way-like period distribution and constrain only the binary fraction $B$ 
in order to distinguish binary velocity variation from measurement error.  
Furthermore, as will become evident shortly, the binary likelihoods become 
computationally expensive to calculate while varying the error parameters if a 
star contains more than one ``long'' time interval ($\gtrsim$ 10 days) between 
measurements.  Thus, any extra measurements occurring at later time intervals 
longer than 10 days will be discarded for the purpose of determining the error 
model parameters, although they will be used later to help determine the best 
possible binary constraints.

In summary, while it may appear cumbersome to apply the binary model twice, the 
benefit of finding more detailed binary constraints \emph{after} determining 
the measurement errors is that ultimately the binary parameters $\mathcal{P}$ 
will be constrained in addition to the binary fraction $B$, and all the 
available measurements will be used to make this determination.

To incorporate binary motion into our likelihood, we need to be able to quickly 
calculate the binary likelihood for a given set of assumed measurement errors, 
which we continually update as the error model parameters are changed. To 
accomplish this, we first run a Monte Carlo simulation to generate the binary 
likelihood in $\log |\Delta v|$ with \emph{zero} measurement error, which we 
denote by $L_b(\log |\Delta v|; \Delta t,M_V)$ where $M_V$ is the absolute 
magnitude of the star and $\Delta t$ is the time interval between measurements.  
The likelihood is generated in $\log |\Delta v|$, rather than $\Delta v$, 
because the binary likelihood becomes singular as $\Delta v \rightarrow 0$. The 
binary likelihood is generated over a table of absolute magnitude values $M_V$ 
and time intervals $\Delta t$.  In practice the long time intervals are 
typically close to a multiple of one year, so for our purposes it is sufficient 
to pick time intervals of 1 year, 2 years, and so on.  For each star with a 
long time interval between measurements, we interpolate in magnitude and choose 
the time interval (in multiples of one year) closest to the actual interval 
between measurements. It should be emphasized that in later sections, we will 
use the exact time intervals between measurements when deriving detailed binary 
constraints; however, for the purpose of deriving measurement errors, 
quantizing the time intervals in this way is less computationally intensive, 
and does not impact the binary modeling or the measurement error results 
significantly.

Now suppose that the star has $n$ velocity measurements, and the long time 
interval is between measurement $k$ and $k+1$. To find the binary likelihood in 
the presence of measurement error, we must find the convolution of the binary 
likelihood and the likelihood from Equation \ref{eq:nonbinary_likelihood_errmodel} 
(representing measurement error) in the velocity difference $\Delta v_k$. By 
switching to an integral over $\log |\Delta v_k'|$, this can be shown to equal

\begin{eqnarray}
\lefteqn{\mathcal{L}_b(\Delta v_i|R_i,c_i,\Delta t_i,M;\mathcal{E})} \nonumber \\
= && \frac{1}{2}\int_{-\infty}^{\infty} L_b(\log |\Delta v_k'|;\Delta t_k,M) \nonumber\\
&& \times  \Big\{\mathcal{L}(\cdots,\Delta v_k - |\Delta 
v_k'|,\cdots|R_i,c_i;\mathcal{E}) + \nonumber \\
&& \mathcal{L}(\cdots,\Delta v_k + |\Delta 
v_k'|,\cdots|R_i,c_i;\mathcal{E})\Big\} d(\log |\Delta v_k'|) \nonumber\\
&&
\end{eqnarray}
where $\mathcal{L}(\Delta v_1,\cdots,\Delta v_{n-1}|R_i,c_i;\mathcal{E})$ is 
the likelihood in Equation \ref{eq:nonbinary_likelihood_errmodel}.  This is our 
binary likelihood expressed in terms of the error model parameters, which must 
be calculated by performing the integral every time the  error model parameters 
are varied. By now it should be evident why only one long time interval ($>$ 10 
days) between measurements is allowed: multiple long time intervals would 
require multiple integrations to account for binary variability over each 
interval, and this would be prohibitively expensive to compute while varying 
the error parameters.

Finally, for each star with a time interval longer than a week, we express the 
likelihood in terms of the binary fraction $B$ as

\begin{eqnarray}
\lefteqn{\mathcal{L}(\Delta v_i|R_i,c_i,\Delta t_i,M;B,\mathcal{E}) } \nonumber\\
& = & (1-B)\mathcal{L}_{nb}(\Delta v_i|R_i,c_i;\mathcal{E}) + B\mathcal{L}_b(\Delta v_i|R_i,c_i,\Delta t_i,M;\mathcal{E}) \nonumber\\
&&
\end{eqnarray}

We shall therefore constrain the four error model parameters $\mathcal{E}$ and 
binary fraction $B$ simultaneously. We emphasize again that once the 
measurement errors have been determined by this procedure, we shall go back and 
find more precise binary constraints by including all the measurements. The 
maximum likelihood error model parameters obtained by this procedure will be 
used to determine the measurement errors for each star before proceeding with 
the full binary analysis.

\begin{figure*}
	\centering
	\subfigure[Fornax]
	{
		\includegraphics[height=0.47\hsize,width=0.35\hsize,angle=-90]{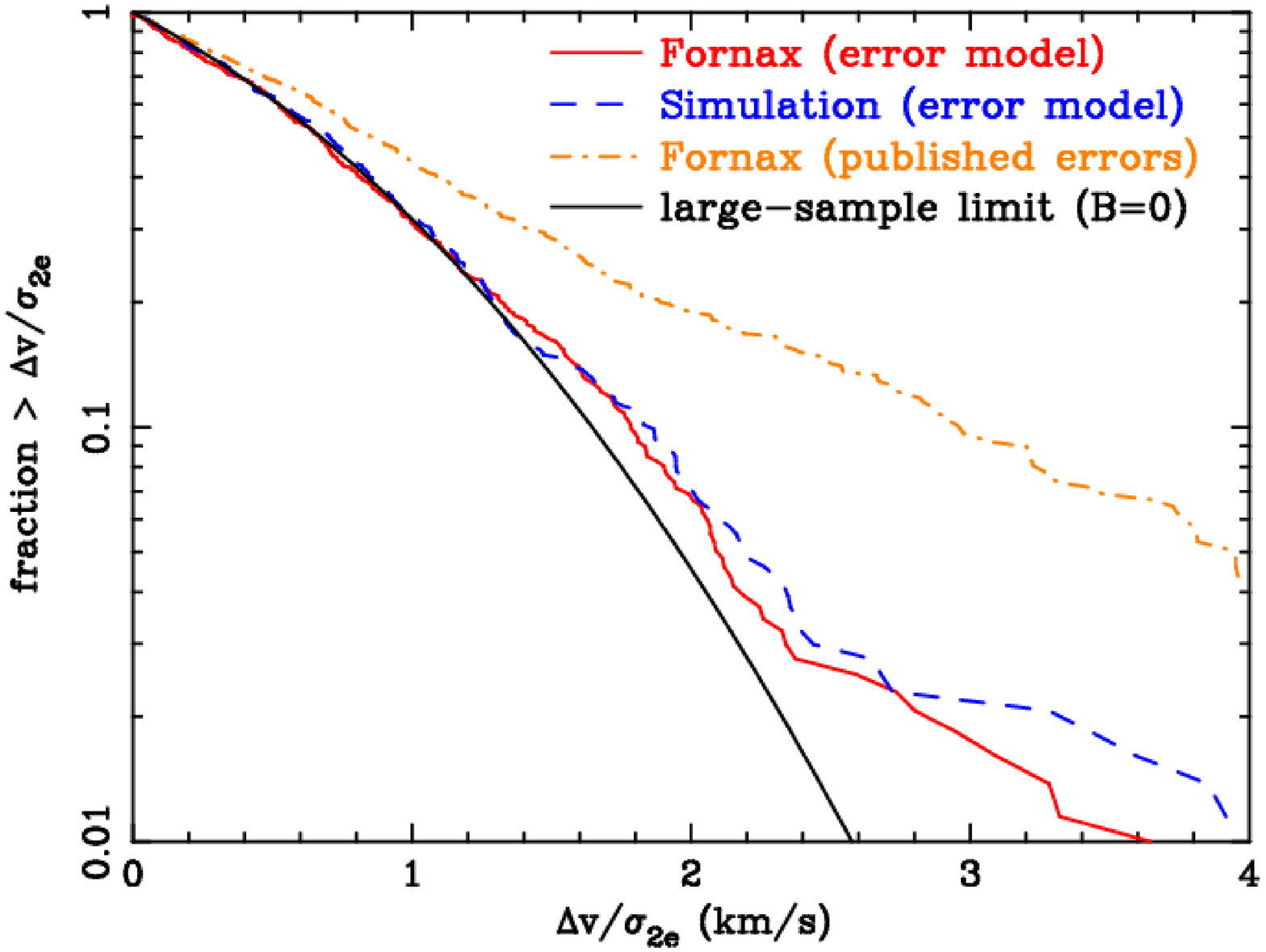}
		\label{fig:dverrchists_fornax}
	}
	\subfigure[Carina]
	{
		\includegraphics[height=0.47\hsize,width=0.35\hsize,angle=-90]{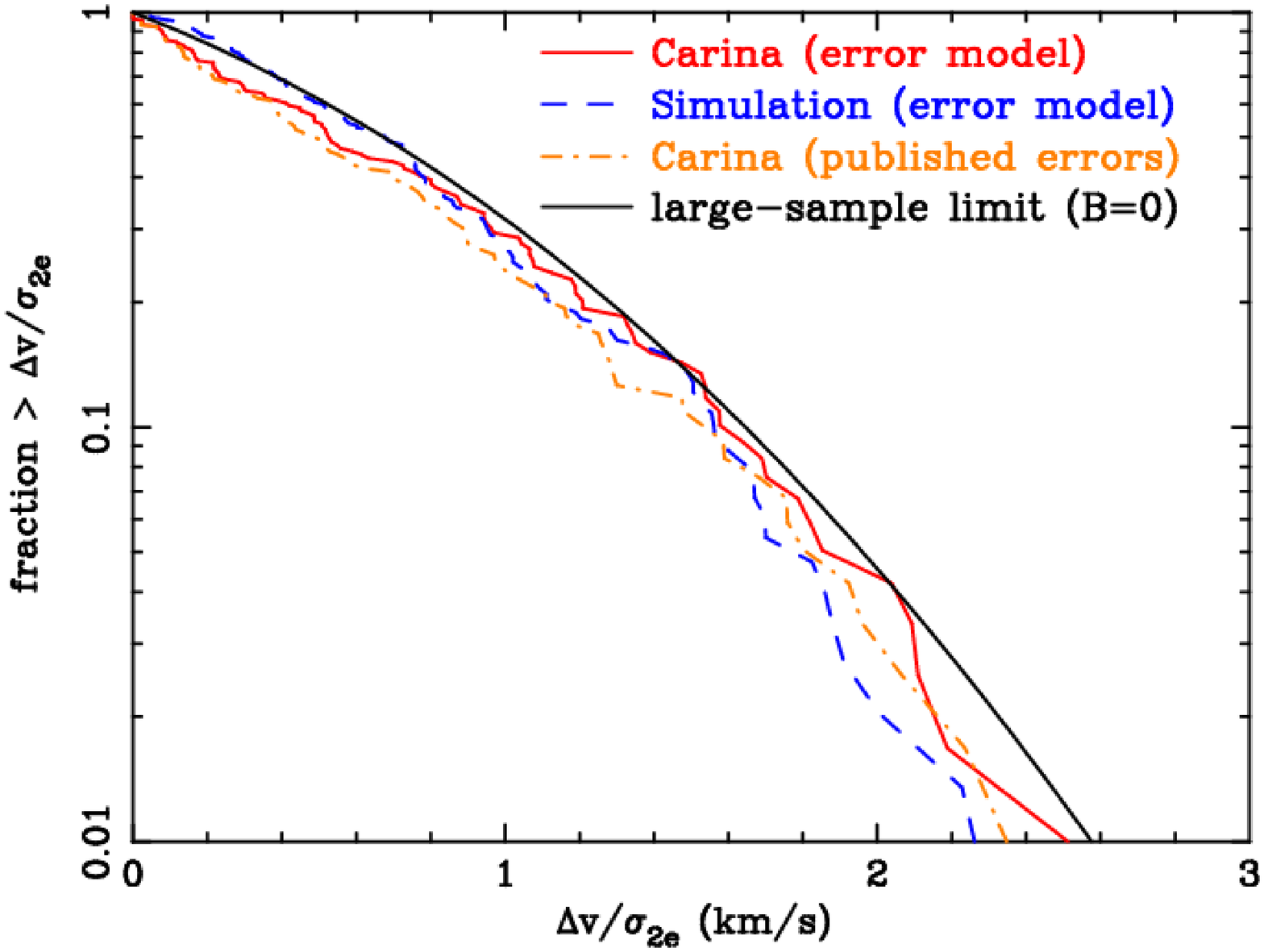}
		\label{fig:dverrchists_carina}
	}
	\subfigure[Sculptor]
	{
		\includegraphics[height=0.47\hsize,width=0.35\hsize,angle=-90]{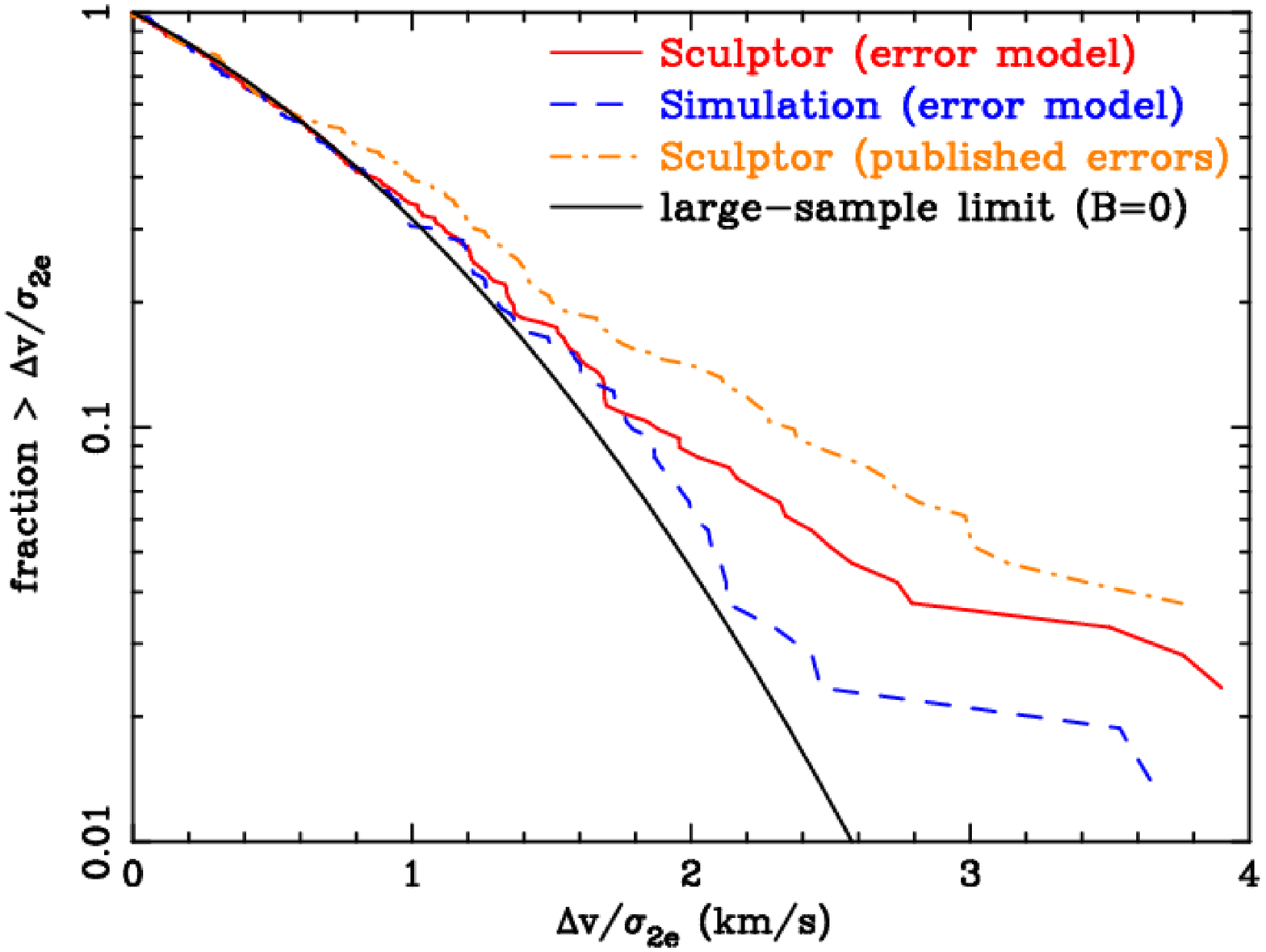}
		\label{fig:dverrchists_sculptor}
	}
	\subfigure[Sextans]
	{
		\includegraphics[height=0.47\hsize,width=0.35\hsize,angle=-90]{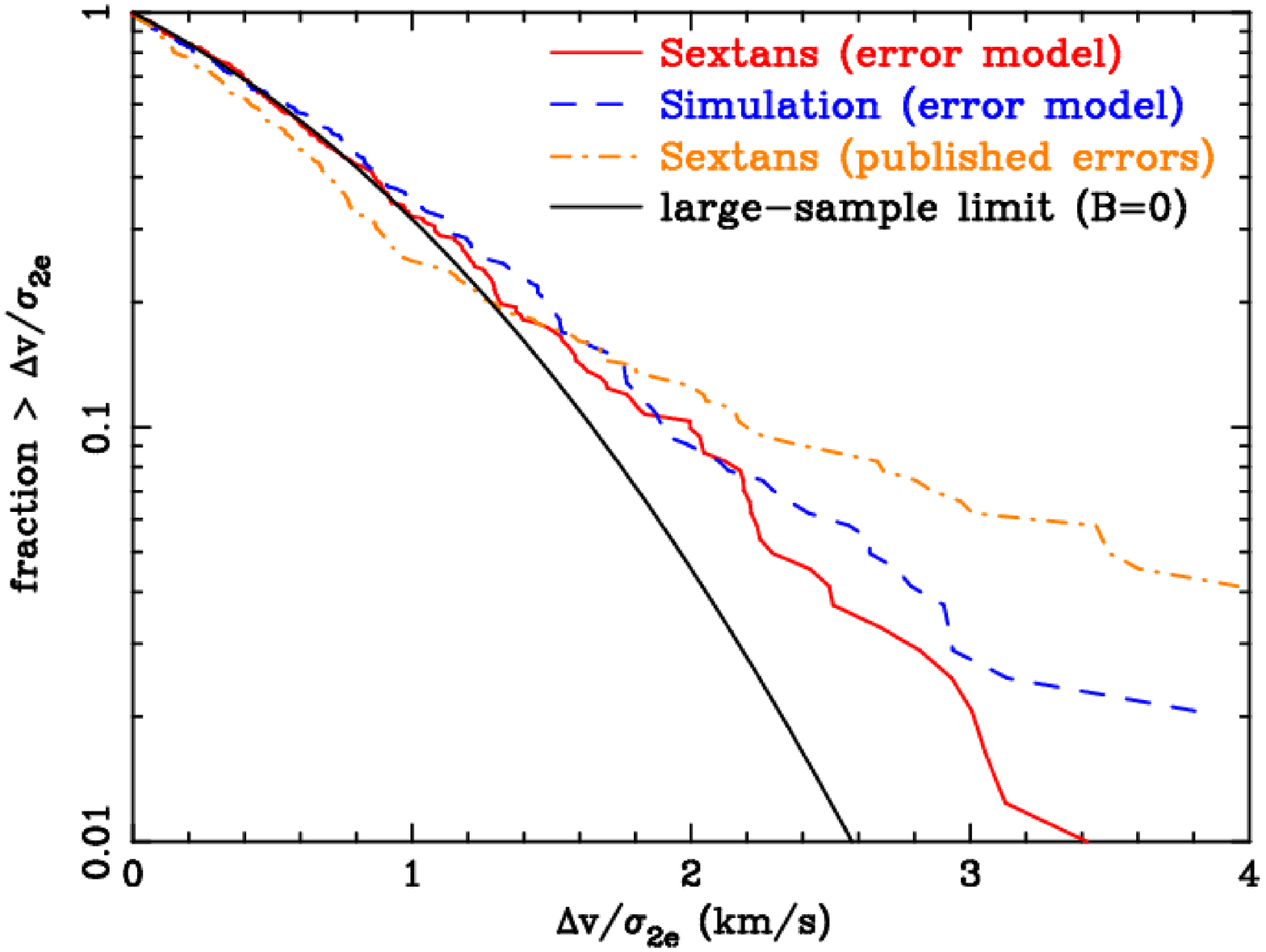}
		\label{fig:dverrchists_sextans}
	}
	\caption{Cumulative distribution of $\Delta v/\sigma_{2e}$ for each galaxy 
in the Magellan/MMFS sample of \cite{walker2009}.  Here, $\Delta v = |v_2-v_1|$ 
is the difference between two successive line-of-sight velocity measurements 
and $\sigma_{2e}$ is the measurement error in $\Delta v$, given by 
$\sigma_{2e}^2 = \sigma_1^2 + \sigma_2^2$ where $\sigma_1$, $\sigma_2$ are the 
individual errors on each measurement. We plot the distribution using 
measurement errors determined from our error model (light solid line). To check 
that the data can indeed be reproduced by our derived error model parameters, 
we plot a distribution for simulated data using the same number of stars, 
derived measurement errors, and epochs as in the real data (dashed line).  For 
comparison, the same distributions are plotted using the published measurement 
errors from \cite{walker2009} (dot-dashed line). The expected distribution in 
the limit of a large sample without binaries is also plotted for reference 
(solid dark line).}
\label{fig:dverrchists_data}
\end{figure*}

\section{Application of error model to data}\label{sec:error_model_data}

We now apply the error model described in the previous section to derive 
measurement errors in the Carina, Fornax, Sculptor, and Sextans dSph galaxies.  
Before plunging into the analysis, we note that there are potential sources of 
non-Gaussian velocity variability besides binary orbital motion. The most 
common of these are due to false peak selection in the spectrum 
cross-correlation function, which occurs at low Tonry-Davis $R$-values and low 
signal-to-noise ratios.  In \cite{walker2009}, measurements with $R$-values 
less than 4 are discarded for this reason. We take the additional step of 
discarding measurements with low S/N, which can also exhibit 
non-Gaussian error--this is apparent from the fact that the distribution of 
velocity changes $\Delta v$ is noticeably non-Gaussian with the inclusion of 
low S/N measurements, even over timescales too small ($\sim$ a few days) for 
binary velocity changes to be observed. On the other hand, the derived 
measurement errors are not significantly affected for S/N thresholds greater 
than $\sim$ 1, except for cuts well above 2 when the sample size is reduced 
significantly.  For this reason, in the following analyses we shall discard 
measurements with S/N less than 1.2.  While this S/N 
threshold might appear somewhat low, it should be borne in mind that the 
original $R$-value cut in \cite{walker2009} already removed many low S/N 
measurements with poor cross-correlation fits, since these are generally 
correlated with low $R$-values. However, there are a large number of 
measurements with S/N between 1 and 2 that have high $R$-values (up to $\sim 
30$), and these are very unlikely to be spurious measurements due to noise.  
Thus, we find that a S/N cut at 1.2 is sufficient to eliminate most 
non-Gaussian error without sacrificing an unnecessarily large number of genuine 
measurements. We do find, however, that measurements using the red channel with 
$R$-values less than 7 correlate strongly with low S/N ($\lesssim$ 
1), due to the red channel's lower spectral resolution compared to the blue 
channel.  Therefore, as a further precaution these measurements in the red 
channel are also discarded to avoid non-Gaussian errors.

When generating the binary likelihoods, the star's magnitude is used to infer 
its mass and stellar radius under the assumption that it lies on the red giant 
branch (see \citealt{minor2010} for details); thus it is essential to exclude 
horizontal branch stars from each sample. In the Carina and Sculptor samples we 
discard stars with magnitudes $m_V > 20.3$ to ensure that horizontal branch 
stars are not included. The Fornax and Sculptor samples do not extend to 
sufficiently faint magnitudes to include horizontal branch stars, so no 
magnitude cuts are necessary in those samples.

As outlined in Section \ref{sec:error_model}, our error model presently allows 
only one ``long'' time interval ($> 10$ days) between measurements---any 
further measurements occurring at long time intervals are discarded, although 
they will be included later when deriving detailed binary constraints in 
Sections \ref{sec:binary_fraction} and \ref{sec:period_distribution}.  After 
making the aforementioned S/N, $R$-value, velocity and magnitude 
cuts in the Fornax sample, there remain 458 stars with repeat measurements, the 
largest of any galaxy in the survey. Of these, however, only 201 stars have 
time intervals of 1 year or longer between any pair of measurements.  Many of 
the latter subset have measurements at three epochs, with time intervals of 
approximately a week and one or two years between measurements.  Since none of 
the stars have multiple time intervals greater than 10 days, no further 
measurements were discarded for the purpose of determining measurement errors 
in Fornax.

After applying our error model to the data, the results for the Fornax dwarf 
galaxy are plotted in Figure \ref{fig:dverrchists_fornax}. For all repeat 
measurements in the sample, we plot the cumulative distribution of $\Delta 
v/\sigma_{2e}$, where $\Delta v = |v_2-v_1|$ is the difference between two 
successive line-of-sight velocity measurements and $\sigma_{2e}$ is the 
measurement error in $\Delta v$, given by $\sigma_{2e}^2 = \sigma_1^2 + 
\sigma_2^2$ where $\sigma_1$, $\sigma_2$ are the individual errors on each 
velocity measurement.  If the velocity changes are only due to Gaussian 
measurement errors, in the limit of a large sample this distribution should 
approach the complimentary error function $\textrm{erfc}\left[\Delta 
v/\sqrt{2}\sigma_{2e}\right]$ (solid dark line). 

\begin{figure*}
	\centering
	\subfigure[200 stars]
	{
		\includegraphics[height=0.47\hsize,width=0.35\hsize,angle=-90]{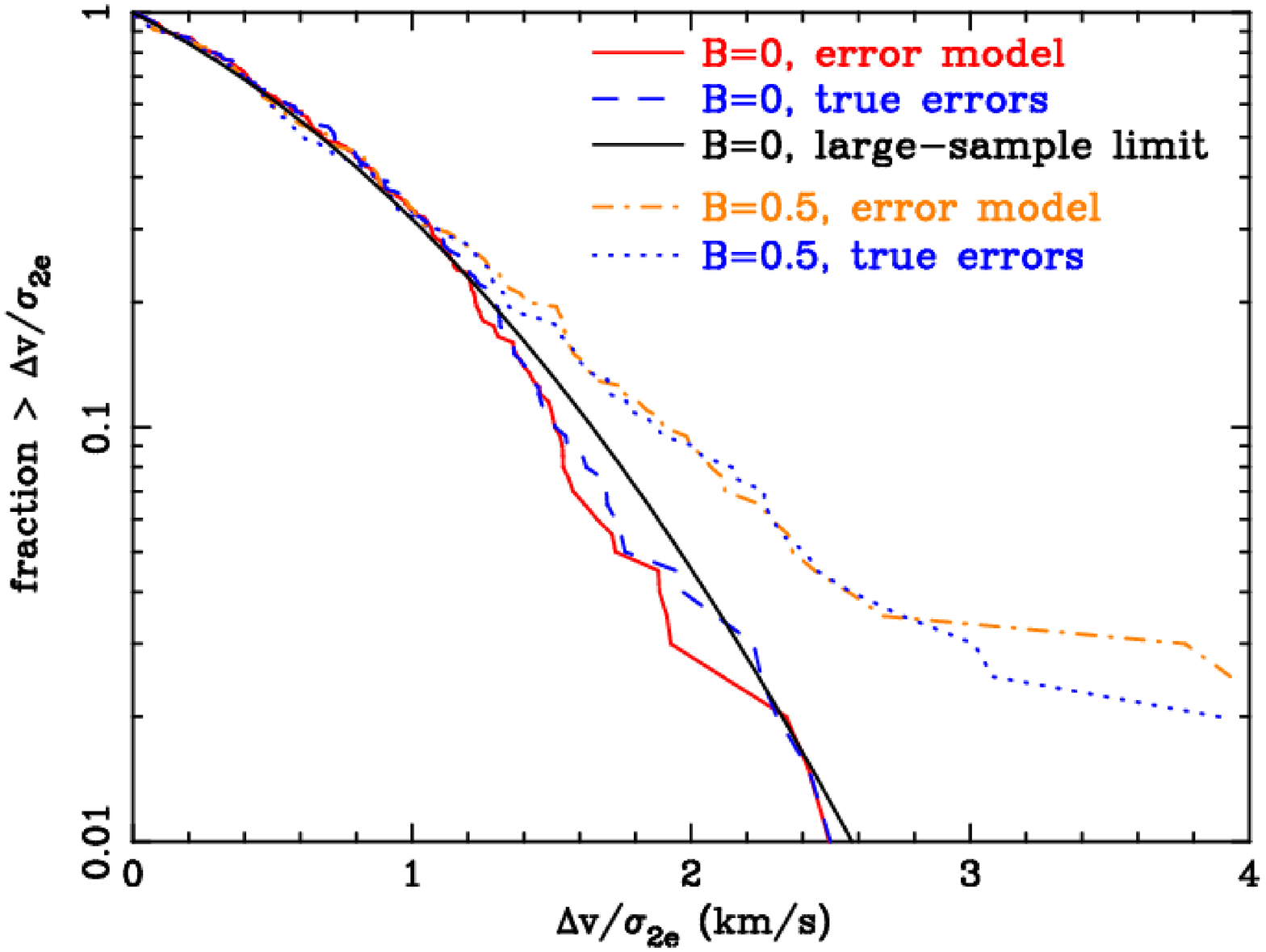}
		\label{fig:dverrchists_2epoch_a}
	}
	\subfigure[200 stars, 2nd realization]
	{
		\includegraphics[height=0.47\hsize,width=0.35\hsize,angle=-90]{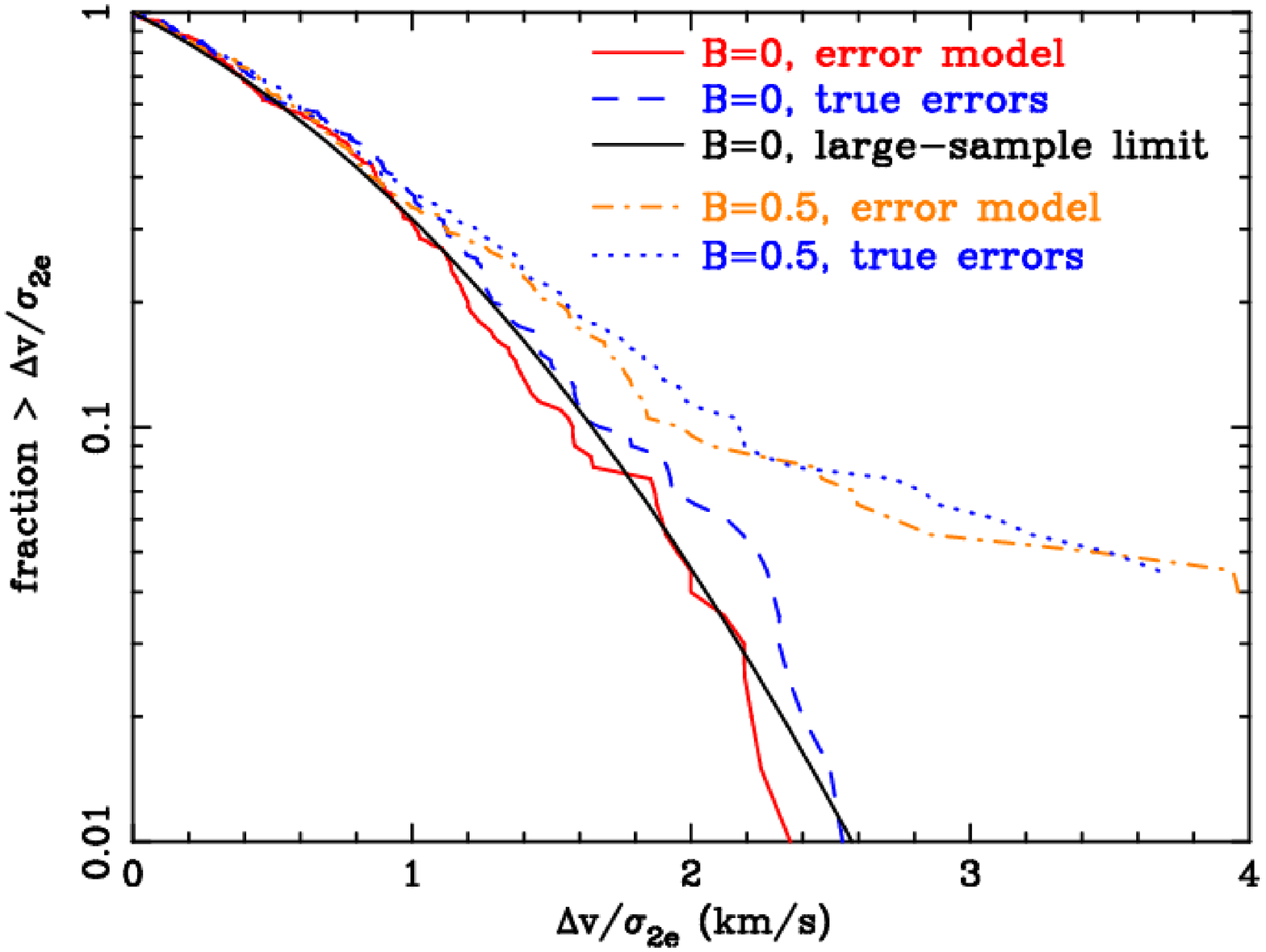}
		\label{fig:dverrchists_2epoch_b}
	}
	\subfigure[500 stars]
	{
		\includegraphics[height=0.47\hsize,width=0.35\hsize,angle=-90]{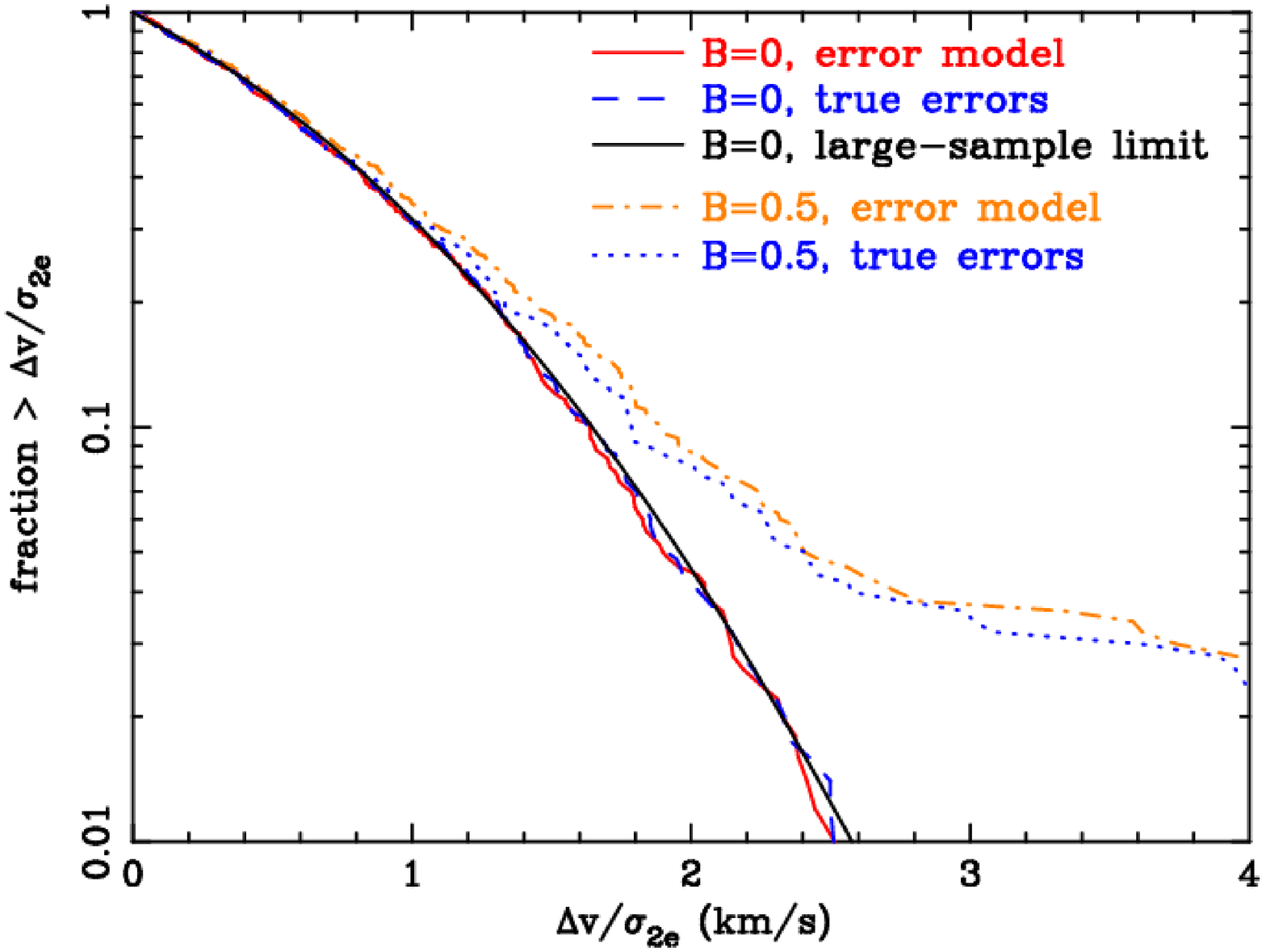}
		\label{fig:dverrchists_2epoch_c}
	}
	\subfigure[500 stars, 2nd realization]
	{
		\includegraphics[height=0.47\hsize,width=0.35\hsize,angle=-90]{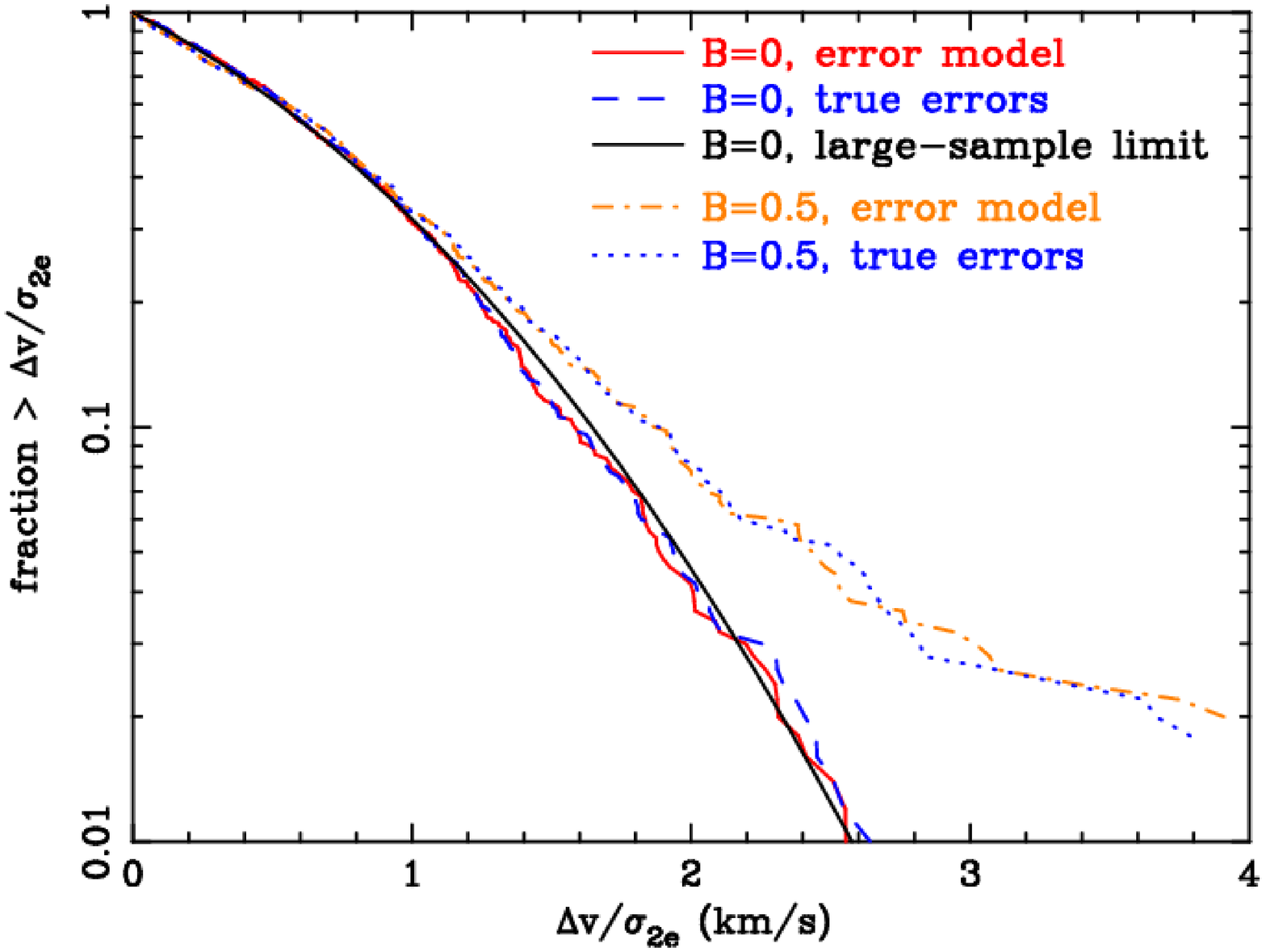}
		\label{fig:dverrchists_2epoch_d}
	}
	\caption{Cumulative distribution of $\Delta v/\sigma_{2e}$ for simulated 
data sets with 2-epoch measurements, with a time interval of 1 year between 
measurements. Shown are two random realizations of data sets with 200 
and 500 stars, respectively, where the distribution of measurement errors is 
similar to those in the data from \cite{walker2009} and the absolute magnitude 
limit $M_{V,lim} = 1$. In each figure we plot a distribution for a sample with 
no binaries, using measurement errors derived by applying our error model to 
the simulated data (light solid line), and a distribution for the same dataset 
using the ``true'' measurement errors of the simulation (dashed line); for 
comparison, we also plot the theoretical expected distribution in the limit of 
a large sample (solid dark line). We also plot distributions for a sample with 
a binary fraction of $B=0.5$ and a Milky Way-like period distribution, again 
using the derived measurement errors from the error model including the effect 
of binary orbital motion (dot-dashed line) and true errors (dotted line) from 
the simulation. In all cases the distributions using the true errors are 
reproduced well by the model.}
\label{fig:dverrchists_2epoch}
\end{figure*}

From the figure it is evident that Gaussian measurement errors are 
well-reproduced in Fornax for $\Delta v \lesssim 1.2\sigma_{2e}$, comprising 
$\sim$80\% of the stars in the sample, while binary orbital motion associated 
with large velocity changes becomes manifest above this threshold.  The maximum 
likelihood binary fraction in Fornax is $B \approx 0.5$, while the maximum 
likelihood error parameters are $\mathcal{E} = \{ \alpha_{red}=36.9$, 
$\alpha_{blue}=55.2$, $x_{red}=1.06$ and $x_{blue}=1.37\}$.  To test this, we 
simulated a dataset with the same derived measurement errors, epochs, and 
magnitudes as in the real data, and assigned new velocity values for each 
measurement assuming a star population with binary fraction $B=0.5$.  The 
distribution for a typical realization is plotted in Figure 
\ref{fig:dverrchists_fornax} (dashed line), and is quite consistent with the 
data given the scatter in samples of this size.  By comparison, we plot the 
distribution using the published measurement errors from \cite{walker2009}, 
from which it is evident that the measurement errors are underestimated even 
for small velocity variations. This discrepancy may be not only due to 
binaries, but also the presence of low S/N measurements biasing the 
error model. We thus conclude that the published measurement errors for the 
Fornax dSph are significantly underestimated; in fact the median measurement 
error in the sample using the published errors is 1.1 km~s$^{-1}$, while the median 
measurement error from our model is 1.7 km~s$^{-1}$, differing by a factor of 
$\approx$ 55\%. The median measurement errors in each galaxy are listed in 
Table \ref{tab:median_errors}.

\begin{table}
\centering
\begin{tabular}{|l|c|c|}
\hline
Galaxy & Median error (km~s$^{-1}$) & Median error, published\\
\hline
Carina & 2.3 & 2.6\\
Fornax & 1.7 & 1.1\\
Sculptor & 2.1 & 1.7\\
Sextans & 2.6 & 2.6\\
\hline
\end{tabular}
\caption{Median measurement error comparison}
\label{tab:median_errors}
\end{table}

In the case of the Carina and Sextans samples, contamination by foreground 
Milky Way stars is a major concern because of the relatively low surface 
brightness of these galaxies. In addition, many stars selected for repeat 
measurements have velocities more than $3\sigma$ away from the galaxy's 
systemic velocity, where $\sigma$ is the galaxy's velocity dispersion. Since 
binary velocity variations greater than $50$ km~s$^{-1}$ are extremely 
unlikely, noting that each galaxy has a velocity dispersion of order $\sim 10$ 
km~s$^{-1}$, we therefore discard stars with velocities $> 60$ km~s$^{-1}$ away from 
the galaxy's systemic velocity as probable nonmember stars. (It should be noted 
that a small subset of these stars do indeed show velocity variations beyond 
the measurement error, but with an amplitude significantly smaller than 50 
km~s$^{-1}$.  Given that nearly all have high metallicities, such stars are most 
likely Milky Way binaries.) In the Carina sample, there are 257 stars with 
repeat measurements; after making the rough velocity cut described above, 
together with a magnitude cut at $m_V=20.3$ to get rid of horizontal branch 
stars, only 107 stars remain in the sample.  Similarly, the Sextans sample 
contains 203 stars with repeat measurements, among which only 134 remain after 
making velocity and magnitude cuts. By contrast, the Sculptor sample contains 
198 stars with repeat measurements, of which 190 remain after velocity and 
magnitude cuts, indicating relatively little contamination by nonmember stars.

Applying our error model to the remaining sample in the Carina dSph, we obtain 
an extraordinarily low best-fit binary fraction, $B=0.05$, the value of which 
will be refined in Section \ref{sec:binary_fraction}.  Using our derived 
measurement errors, the resulting distributions in $\Delta v/\sigma_{2e}$ are 
plotted in Figure \ref{fig:dverrchists_carina}.  Because of the relatively 
small size of the sample, the distribution can differ significantly from the 
expected large-sample limit, but the absence of binary velocity variation is 
immediately apparent in the figure (we will return to this in detail in Section 
\ref{sec:binary_fraction}). To test our model, we again simulate a dataset with 
the same derived measurement errors, epochs, and magnitudes as the real data, 
and assign new velocity values for each measurement assuming our best-fit 
binary fraction for Carina. Two random realizations are plotted; although the 
distributions can vary considerably because of the small sample size, we see 
that the actual distribution using our measurement errors is indeed consistent 
with the model.

Finally, we apply our error model to the Sculptor and Sextans dSph samples.  
After discarding measurements in the red channel with low $R$-values, we find 
that the remaining Sextans sample contains only 16 stars with repeat 
measurements in the red channel, considerably less than the other galaxies in 
the sample. As this is an insufficiently small number of repeat measurements to 
derive reliable measurement errors for the red channel, we discard the red 
channel measurements in Sextans altogether for the remainder of this work.  
After making the final cuts, we plot distributions in $\Delta v/\sigma_{2e}$ 
for Sculptor and Sextans in Figures \ref{fig:dverrchists_sculptor} and 
\ref{fig:dverrchists_sextans} respectively.  For each galaxy, we plot 
distributions using our derived measurement errors, along with distributions 
using the published measurement errors. From the figure it is evident that for 
both galaxies, the published data underestimates the measurement error for 
large velocity differences, while in Sextans the error is overestimated for 
small velocity changes.  By contrast, the Gaussian error is well-accounted for 
in our derived measurement errors, with binary velocity variation showing up 
for velocity changes greater than $\approx$ 1.4 km~s$^{-1}$.

While the principle goal of our analysis in this section has been to derive 
measurement errors, a rough estimate of the binary fraction in each galaxy has 
been derived simultaneously. However, the binary constraints obtained in this 
section are only approximately valid, for several reasons: first, to save time 
during the likelihood calculation, for each star with a long time interval ($> 
10$ days) between measurements, the long time interval was rounded to a 
multiple of 1 year---thus the binary likelihoods were generated for multiples 
of 1 year without the need for additional interpolation.  For some stars the 
rounding error is as long as three months, so the binary likelihood is only 
approximately accurate in those cases.  More importantly, as already discussed, 
certain repeat measurements were discarded if the star already contained 
multi-epoch measurements with one long time interval. Some of those discarded 
measurements may indicate binary velocity variation and thus are important for 
finding binary constraints.  Finally, by inferring the binary fraction, we have 
assumed thus far that the distribution of orbital periods in each galaxy is 
identical to that of Milky Way field binaries, which may not be the case.  
Ultimately, it would be preferable to constrain not only the binary fraction, 
but also the period distribution parameters in each galaxy. The full binary 
analysis will be performed in Sections \ref{sec:binary_fraction} and 
\ref{sec:period_distribution}.

\section{Test of error model on simulated data}\label{sec:test_error_model}

To verify that we have reproduced the Gaussian measurement errors well, we 
shall apply our error model to a series of simulated datasets with a 
distribution of Gaussian measurement errors similar to that obtained in the 
Magellan/MMFS survey of \cite{walker2009}. This is accomplished by drawing a 
random Tonry-Davis $R$-value for each measurement from a Gaussian distribution 
whose mean and spread are similar to those observed in the real data, while the 
spectrograph channel used to obtain each measurement is chosen to be either red 
or blue with equal probability. The ``true'' error model parameters are chosen 
to be identical to those obtained from the Fornax sample (which have been 
obtained in Section \ref{sec:error_model_data}). First we consider samples with 
measurements at two epochs, separated a year apart, with sample sizes of 200 
and 500 stars.  In each case, we simulate a sample with a binary fraction of 
zero and apply our error model without correcting for binaries. We also 
simulate a sample with a binary fraction $B=0.5$ and apply our error model with 
the binary correction. In all simulations, an absolute magnitude limit $M_V = 
1$ is assumed.

The results are displayed in Figure \ref{fig:dverrchists_2epoch}. As in the 
previous section, we plot the cumulative distribution of $\Delta 
v/\sigma_{2e}$, where $\Delta v = |v_2-v_1|$ is the difference between two 
successive line-of-sight velocity measurements and $\sigma_{2e} = \sigma_1^2 + 
\sigma_2^2$ where $\sigma_1$, $\sigma_2$ are the individual errors on each 
velocity measurement.  In Figures \ref{fig:dverrchists_2epoch_a} and 
\ref{fig:dverrchists_2epoch_b} we consider two typical random realizations of a 
200-star sample, where each star has measurements at two epochs separated by a 
year, and plot distributions using the ``true'' measurement errors for each 
simulation, as well as the derived measurement errors by applying our error 
model to the simulation. In each realization, the distribution using the 
derived errors follows the distribution using the ``true'' errors for both the 
$B=0$ and $B=0.5$ populations; the percent difference between the derived and 
true errors is smaller than 10\% for at least $\approx$ 80\% of the stars in 
the sample, and smaller than 30\% for $\approx$ 95\% of the stars in the 
sample.  The Gaussian measurement errors are thus reasonably well-reproduced by 
the error model for most stars in the sample.  For velocity changes $\Delta v / 
\sigma_{2e} < 1$, the distributions follow the large-sample limit, indicating 
the distribution of velocities below this threshold is Gaussian and binary 
behavior is not distinguishable from measurement error for such small velocity 
changes.  As can be seen from Figures \ref{fig:dverrchists_2epoch_c} and 
\ref{fig:dverrchists_2epoch_d}, the measurement errors are very well-reproduced 
in a 500-star sample regardless of the realization---in both cases shown, the 
percent difference between the derived and true errors is smaller than 10\% for 
at least 97\% of the stars in the sample. Since the distributions are 
cumulative, if a few measurement errors are reproduced incorrectly, the 
distribution using the derived errors can be offset from the true distribution; 
note however that in Figure \ref{fig:dverrchists_2epoch_c} the distribution for 
$B=0.5$ parallels the true distribution, indicating most of the measurement 
errors are indeed accurate.

\begin{figure}
	\includegraphics[height=1.0\hsize,angle=-90]{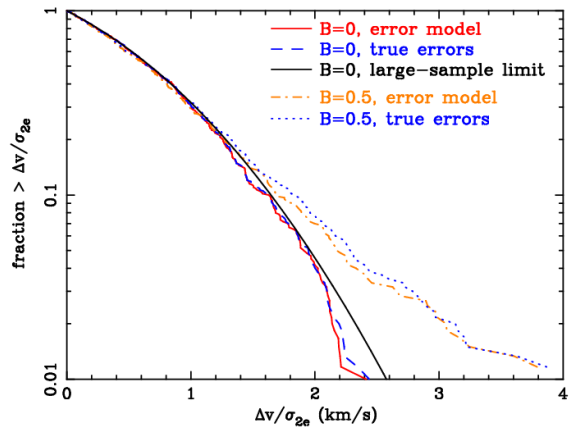}
	\caption{Cumulative distribution of $\Delta v/\sigma_{2e}$ for simulated 
data sets of 200 stars with 4-epoch measurements, where the first time interval 
between measurements is 1 year, and the remaining time intervals is 1 day.  
Plotted curves are identical to those in Figure \ref{fig:dverrchists_2epoch}.  
Note that the measurement errors are better reproduced compared to a similar 
sample with measurements at only 2 epochs (compare 
Figures \ref{fig:dverrchists_2epoch_a} and \ref{fig:dverrchists_2epoch_b}), 
although the binary fraction is not any better constrained because only one 
time interval is long enough for binary variability to be observed.}
\label{fig:dverrchists_gt2ep}
\end{figure}

As a final sanity check, we verify that our error model works for a larger 
number of epochs. In Figure \ref{fig:dverrchists_gt2ep} we consider a sample of 
200 stars with measurements at four epochs, where the first time interval 
between measurements is equal to 1 year, and the remaining time intervals is 
equal to 1 day. Since time intervals shorter than a week are too short for 
binary orbital motion to be evident, we expect the additional epochs to aid in 
constraining the measurement errors, although without helping to constrain the 
binary fraction directly. We find that applying our error model to the 4-epoch 
sample reproduces the measurement errors very well, although the sample with 
$B=0.5$ is not as well reproduced as in the 500-star sample because the binary 
fraction is not as well constrained.

\section{Binary fraction in Carina, Fornax, Sculptor, and Sextans dSph 
galaxies}\label{sec:binary_fraction}

\begin{figure*}
	\centering
	\subfigure[Fornax, Sculptor]
	{
		\includegraphics[height=0.44\hsize,width=0.31\hsize,angle=-90]{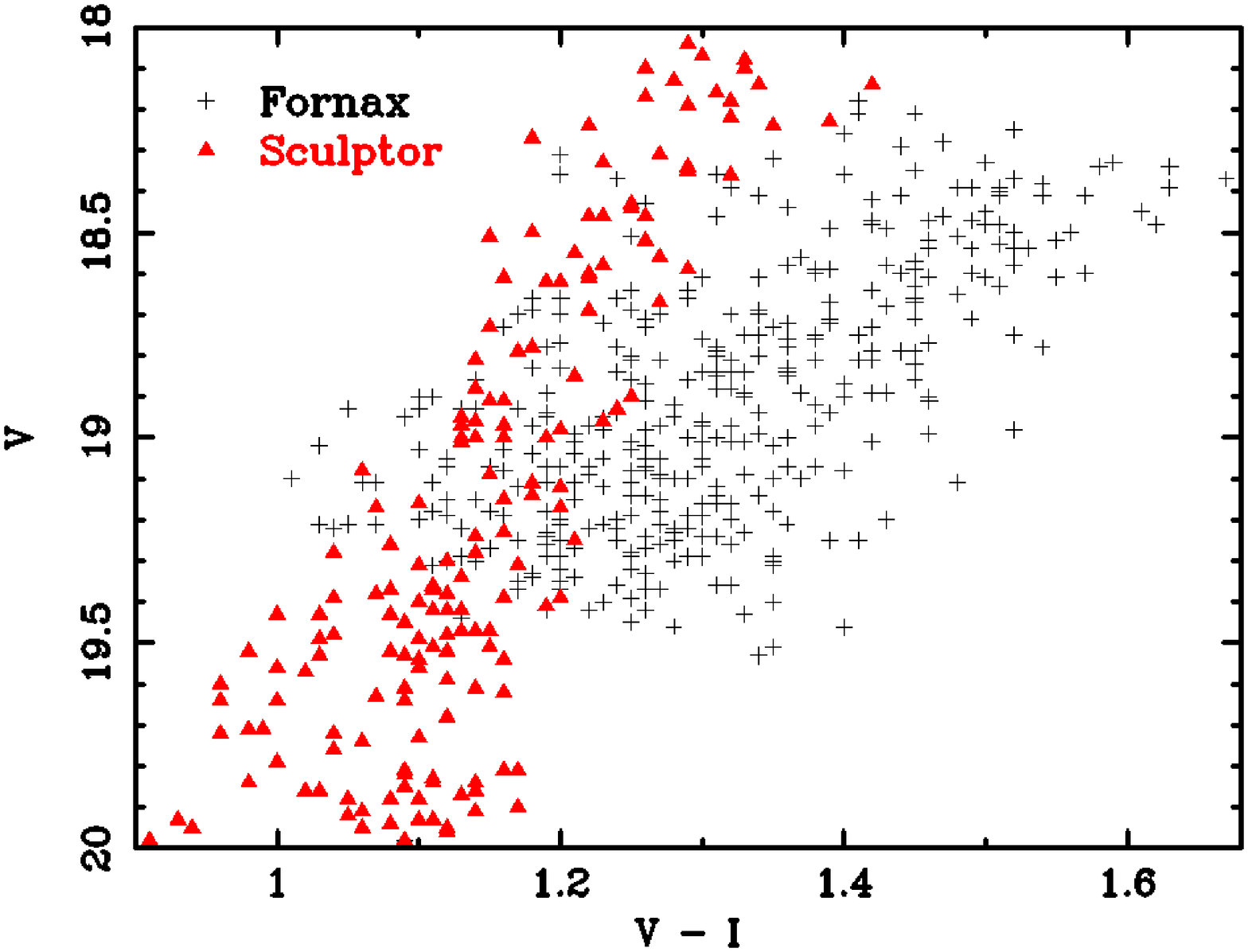}
	}
	\subfigure[Carina, Sextans]
	{
		\includegraphics[height=0.44\hsize,width=0.31\hsize,angle=-90]{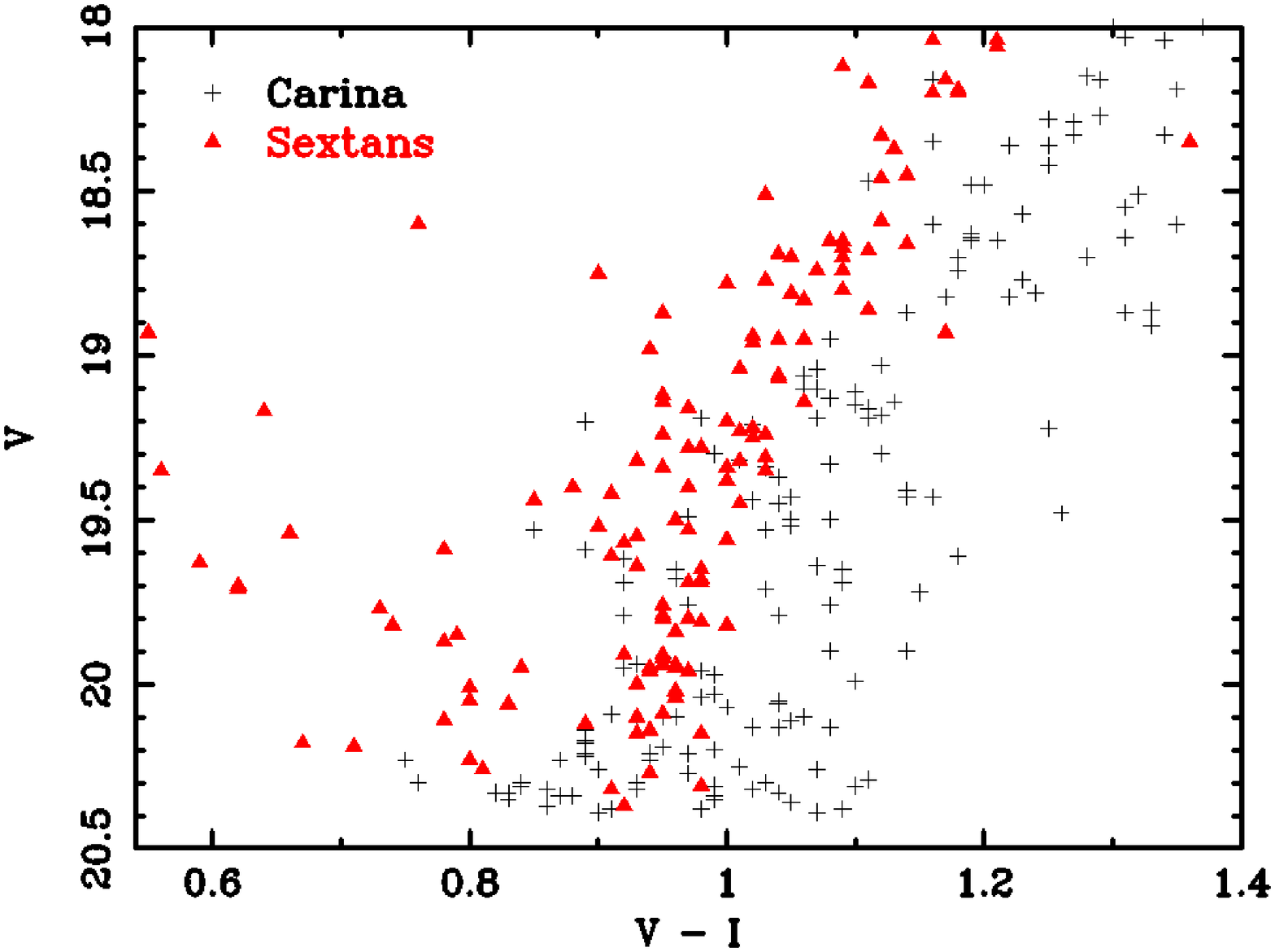}
	}
	\caption{Color-magnitude diagrams for the sample in each galaxy after 
	applying our selection criteria. Only stars with measurements at more than 
	two epochs are included in the final sample (measurements spaced less than 
	10 days apart are averaged together and considered a single epoch).  In 
	Carina and Sextans, stars with $m_V > 20.4$ are excluded, and additional 
	color-magnitude cuts are applied to remove horizontal branch stars. For all 
	samples, measurements with S/N less than 1.2 are dropped.}
	\label{fig:cm_diagrams}
\end{figure*}

\subsection{Sample selection criteria; contamination by Milky Way stars}

We now adopt a Bayesian approach to infer properties of the binary populations 
in each galaxy. For simplicity, in this section we will assume the distribution 
of binary periods in each galaxy is identical to that of Milky Way field 
binaries (with a mean period of $\sim$180 years and log-spread of periods equal 
to 2.3; see \citealt{duquennoy1991}), an assumption that will be relaxed in 
Section \ref{sec:period_distribution}.

For reasons discussed in Section \ref{sec:error_model_data}, we make a 
magnitude cut to exclude horizontal branch stars in the Carina and Sextans 
galaxies.  Because these galaxies have relatively small member samples, 
however, it would be preferable to make a more specific color-magnitude cut to 
exclude the horizontal branch without sacrificing too many red giant branch stars.  To 
accomplish this, we cut out stars with $m_V > 20.4$ for both galaxies, but make 
an additional color-magnitude cut to exclude the bright end of the horizontal 
branch. For Carina, stars with $m_V > 20.35$ and color $V-I < 0.85$ are 
probable red clump horizontal branch stars and are therefore excluded; 
likewise, in Sextans, a similar cut is made for $m_V > 20.3$ and $V-I < 0.8$.  
(For color-magnitude diagrams of the red giant branch and horizontal branch region in each galaxy, we refer the 
reader to \citealt{walker2007}). We also discard measurements with low 
S/N to avoid non-Gaussian errors that may bias the inferred binary 
fraction.  As we will show specifically for Carina in Section 
\ref{sec:carina_low_bf}, our results are essentially unchanged for 
S/N thresholds between 1.2 and 2, except that the constraints 
become weaker for higher thresholds as a greater fraction of stars are being 
removed from the sample.  Therefore, as in Section \ref{sec:error_model_data}, 
we will discard measurements with S/N smaller than 1.2 in the 
calculations that follow. To eliminate obvious nonmember stars, we make a rough 
velocity cut as described in Section \ref{sec:error_model_data} for each 
galaxy.  Finally, measurements that occur within 10 days of each other are 
averaged together (using Equations \ref{eq:average_v} and \ref{eq:sigm}), since 
binary velocity variation cannot be observed on shorter timescales given the 
measurement errors. Only stars with repeat measurements at two or more epochs 
(after this averaging) are included in the final sample.  Color-magnitude 
diagrams for our final sample in each galaxy are shown in Figure 
\ref{fig:cm_diagrams}.

Another potential bias in the inferred binary fraction is contamination by 
foreground Milky Way binary stars. The majority of foreground stars are 
K-dwarfs in the Milky Way disk, many of which can be expected to have binary 
companions.  We can expect larger velocity variations in these binaries 
compared to binaries in the background dSph galaxy, for two reasons: first, 
since the observed nonmember star lies on the main sequence, the secondary star 
may not be significantly dimmer than the primary star. Thus the measured radial 
velocity may be that of the less massive secondary star, for which larger 
velocity amplitudes are expected.  Second, since the observed member stars lie 
on the red giant branch, very close binaries with high velocity amplitudes will 
have been destroyed by Roche-lobe overflow (\citealt{paczynski1971}), whereas 
this is not necessarily the case for the foreground Milky Way stars since the 
observed star lies on the main sequence.  Thus if there is significant 
contamination by Milky Way binaries, we can expect that it will most likely 
bias the inferred binary fraction to a high value.

We can cut down on contamination significantly by performing a rough velocity 
cut as outlined in Section \ref{sec:error_model_data}. In addition, since the 
metallicity distribution is well-determined in each galaxy in terms of the Mg I 
triplet ($\lambda \sim 5170 \AA$) absorption line strength, we can use 
metallicity information to help distinguish between member and non-member 
stars. Following \cite{walker11-09}, we model the distribution of Mg-triplet 
pseudo-equivalent widths (or magnesium strengths) for both the member and 
nonmember stars as a Gaussian, so that the member star population has mean 
magnesium strength $\bar w$ and dispersion in magnesium strength $\sigma_w$; 
likewise, the nonmember stars have similar parameters $\bar w_{MW}$, 
$\sigma_{w,MW}$. The values of each of these parameters characterizing the 
metallicity distribution in each sample are determined in \cite{walker11-09}.  
For each individual star we calculate the average magnesium strength $w$ by 
performing a weighted average over all measurements $w_i$, and likewise we 
calculate the average measurement error in magnesium strength, $e_w$.  Assuming 
Gaussian measurement errors in $w$, we then define the likelihood in magnesium 
strength for member stars as

\begin{equation}
\mathcal{L}_w(w_i|e_{w,i}) = \frac{1}{\sqrt{2\pi(e_w^2 + \sigma_w^2)}}\exp\left[\frac{-(w-\bar w)^2}{2(e_w^2 + \sigma_w^2)}\right]
\end{equation}

The corresponding likelihood for Milky Way stars is identical, except with the 
Milky Way magnesium strength distribution parameters $\bar w_{MW}$, 
$\sigma_{w,MW}$.  For the member stars, we use the same velocity likelihood as 
in Equation \ref{eq:full_binary_likelihood}, while for the nonmember stars we 
use the nonbinary velocity likelihood in Equation 
\ref{eq:nonbinary_likelihood}.  Assuming a fraction $F$ of stars in the sample 
are members, our likelihood can be written

\begin{eqnarray}
\label{eq:full_vw_likelihood}
\lefteqn{
\mathcal{L}(\Delta v_i, w_i|e_{w,i},\sigma_i,t_i,M; B, \mathcal{P})} \\ \nonumber
& = & (1-F)\mathcal{L}_{MW}(\Delta v_i|\sigma_i)\mathcal{L}_{w,MW}(w_i|e_{w,i}) \\ \nonumber
& & + F\mathcal{L}(\Delta v_i|\sigma_i,t_i,M;B,\mathcal{P})\mathcal{L}_w(w_i|e_{w,i})
\end{eqnarray}

Since we are not modeling the binary population of the Milky Way stars, there 
is still the possibility that a binary nonmember star may be counted as a 
member if it exhibits a large velocity change.  However, the stellar 
metallicities in the Sculptor, Sextans, and Carina galaxies are significantly 
lower than that of the nonmember stars, so the metallicity likelihoods will 
significantly reduce the impact of these stars on the inferred binary fraction.  
The Fornax dSph has a metallicity distribution similar to that of the nonmember 
stars, so the method outlined above cannot resolve nonmember contamination in 
Fornax; fortunately however, because of its high surface brightness, the 
fraction of nonmember stars in the Fornax sample is expected to be very low.

\begin{figure}
	\includegraphics[height=1.0\hsize,angle=-90]{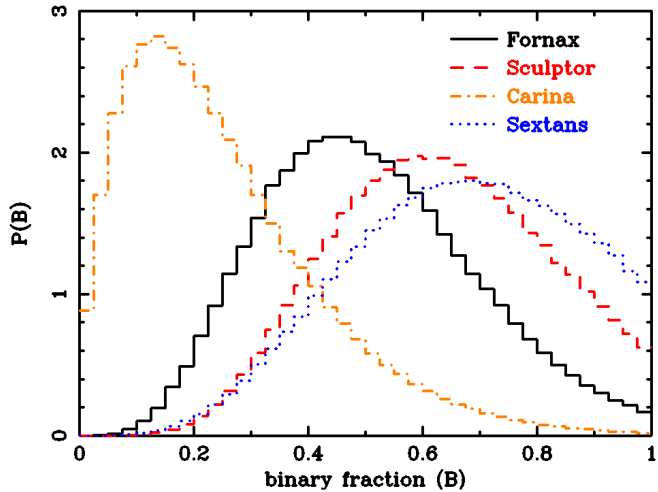}
	\caption{Posterior probability distributions in binary fraction for each of 
the galaxies in the Magellan/MMFS sample of \cite{walker2009}, where the binary 
period distribution is assumed to be identical to that of Milky Way field 
stars. In the Fornax galaxy (solid curve) the most probable inferred binary 
fraction is consistent with Milky Way field binaries, whereas the Carina dwarf 
(dot-dashed curve) has a very low inferred binary fraction, indicating a dearth 
of short-period binaries compared to the other galaxies.}
\label{fig:bfposts_alldwarfs_sn1.2}
\end{figure}

\subsection{Binary fraction constraints} \label{sec:binary_fraction_constraints}

Adopting the likelihood in Equation \ref{eq:full_vw_likelihood}, we use a nested 
sampling routine (\citealt{skilling2004}; \citealt{feroz2009}) to obtain marginal posterior probability distributions in the 
binary fraction $B$ and member fraction $F$ in each galaxy, assuming a flat 
prior in each parameter. With the exception of Sextans, we find the inferred 
member fraction in each galaxy is greater than 0.9 to within 99\% confidence 
limits, indicating our rough velocity cut (discussed in Section 
\ref{sec:error_model_data}) successfully removed most of the Milky Way stars.  
In Sextans, however, the member fraction is constrained to be 
$0.75_{-0.05}^{+0.04}$, indicating contamination may still be an issue, although 
metallicity information can still distinguish between member and nonmember 
stars in most cases due to Sextans' low mean magnesium line strength ($\bar w = 
0.36 \AA$, compared to $\bar w_{MW} = 0.84 \AA$ for the foreground Milky Way 
stars).  Since the member fraction in each galaxy is well-constrained by the 
metallicities in any case, the result is not strongly dependent on our prior in 
the member fraction $F$.

In Figure \ref{fig:bfposts_alldwarfs_sn1.2}, we plot posterior probability 
distributions in the binary fraction $B$ after marginalizing over the member 
fraction $F$. The most probable inferred binary fractions in each galaxy are 
listed in Table \ref{tab:binary_fraction}, where the errors are given by the 
15.87\% and 84.13\% percentiles of the probability distributions (corresponding 
to 1-$\sigma$ error bars in a Gaussian distribution).

\begin{table}[t]
\centering
\begin{tabular}{|l|c|}
\hline
Galaxy & Binary fraction \\
\hline & \\ [-2.1ex]
Carina & $0.14^{+0.28}_{-0.05}$\\
& \\ [-2.1ex]
Fornax & $0.44^{+0.26}_{-0.12}$\\
& \\ [-2.1ex]
Sculptor & $0.59^{+0.24}_{-0.16}$\\
& \\ [-2.1ex]
Sextans & $0.69^{+0.19}_{-0.23}$\\
& \\ [-2.1ex]
Combined & $0.46^{+0.13}_{-0.09}$\\
\hline
\end{tabular}
\caption{Best-fit binary fractions, assuming Milky Way period distribution}
\label{tab:binary_fraction}
\end{table}

To evaluate whether the inferred binary fractions in Figure 
\ref{fig:bfposts_alldwarfs_sn1.2} are consistent with that of Milky Way field 
binaries, we note that binaries with G-dwarf primaries in the solar 
neighborhood are found to have a binary fraction of $\approx$ 0.5 
(\citealt{duquennoy1991}); however, binary fraction is known to be a function 
of mass, with smaller-mass primaries having smaller binary fractions 
(\citealt{fischer1992}, \citealt{raghavan2010}). In the solar neighborhood, 
\cite{raghavan2010} finds that primary masses in the approximate range of 
0.75-1$M_\odot$ correspond to an average binary fraction of $\sim$0.4, while 
1-1.4$M_\odot$ primaries correspond to an average binary fraction of $\sim$0.5.  
The galaxies in this study have multiple star populations with widely varying 
ages, and thus a range of primary masses on the red giant branch. While 
photometry shows the Sculptor and Sextans dSphs have predominantly very old 
star populations with ages $\gtrsim$ 10 Gyr (\citealt{deboer2012_sculptor}, 
\citealt{lee2009}), the Fornax dSph shows a wide spread of stellar ages from 2 
Gyr up to 13 Gyr (\citealt{deboer2012}), while Carina's dominant stellar 
population has ages of 4-7 Gyr (\citealt{hurley-keller1998}). On the red giant 
branch, this corresponds to stellar masses ranging from $0.8 M_\odot$ up to 
approximately $1.2 M_\odot$. We therefore assume an inferred binary fraction in 
the range 0.4-0.6 to be roughly consistent with that of Milky Way field 
binaries.

By this standard, Fornax, Sculptor, and Sextans all have binary fractions 
consistent with Milky Way field binaries to within 68\% confidence limits.  
Carina, by contrast, has a most probable inferred binary fraction of 0.14, and 
its inferred binary fraction is less than 0.29 to within 68\% confidence 
limits, and less than 0.57 to within 95\% confidence limits. It must be borne 
in mind that here we have assumed Carina's period distribution to be similar to 
that of the Milky Way.  More generally, the results imply that Carina is 
largely devoid of binaries with periods ($\lesssim$ 10 years) that would 
produce observable velocity changes over the relevant timescale of a few years.  
As will be made explicit in Section \ref{sec:period_distribution}, this could 
also be explained if the mean period of Carina's binary population is longer 
than that of the Milky Way, or has a smaller spread of periods, rather than 
having a very small binary fraction.  Whatever the reason, Carina's lack of 
short-period binaries appears to be inconsistent with having a binary 
population similar to that of Milky Way field binaries. We will discuss the 
possible implications of this result for star formation in Section 
\ref{sec:discussion}.

\begin{figure}
	\includegraphics[height=1.0\hsize,angle=-90]{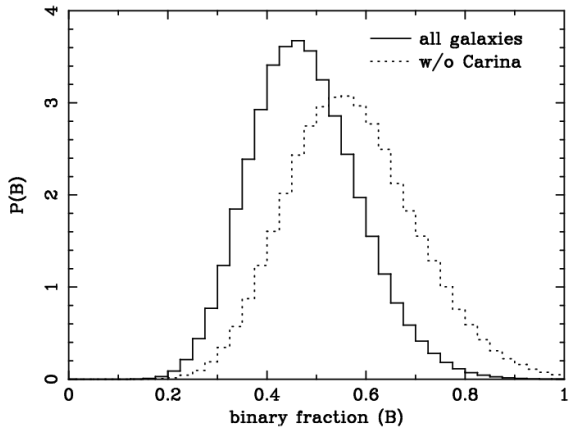}
	\caption{Posterior probability distribution in binary fraction using the 
combined sample of galaxies in the Magellan/MMFS dataset, where the binary 
period distribution is assumed to be identical to that of Milky Way field 
stars. Plotted are the combined sample of all four galaxies (solid curve) and 
the combined sample of Fornax, Sculptor and Sextans (dotted curve) without 
including the Carina dSph which has an anomalously low binary fraction.  The 
most probable binary fraction over the entire combined sample is 
$0.46^{+0.12}_{-0.09}$, consistent with the Milky Way field binary fraction of 
$\approx$ 0.5 for solar-type stars.}
\label{fig:bfposts_combined_dsphs}
\end{figure}

Continuing under the assumption of Milky Way-like period distributions in each 
galaxy, we can ask, what is the binary fraction of the \emph{combined} sample 
of all four galaxies?  There are 621 stars in the combined sample, affording 
much better constraints on binary fraction compared to the sample in each 
galaxy separately. The inferred binary fraction of the combined sample is 
plotted in Figure \ref{fig:bfposts_combined_dsphs} (solid curve), and its 
best-fit value is given in the last entry of Table \ref{tab:binary_fraction}.  
The most probable inferred binary fraction is 0.46, and the 68\% confidence 
interval lies in the range 0.37-0.59, which is consistent with the Milky Way 
field binary fraction $\approx$ 0.5 for solar-type stars. Finally, since Carina 
has an anomalously low apparent binary fraction, we also plot the combined 
sample of Fornax, Sculptor, and Sextans without including the stars from Carina 
(dotted curve). The most inferred probable binary fraction without Carina is 
0.55, with the 68\% confidence interval lying in the range 0.44-0.70, somewhat 
high but still consistent with Milky Way field binaries.

\subsection{Low binary fraction in Carina dSph?} \label{sec:carina_low_bf}

The apparent shortage of short-period binaries in the Carina dSph cannot be 
easily explained by sources of bias in the inferred binary fraction. Systematic 
errors such as contamination by foreground Milky Way stars, non-Gaussian errors 
due to low S/N or other factors, or the presence of variable stars 
all tend to increase the non-Gaussian velocity variability of the stars, and 
thus would \emph{inflate} the inferred binary fraction, rather than reduce it.  
Indeed, such biases might be a factor in the somewhat high inferred binary 
fractions of Sextans and Sculptor. In particular, contamination by Milky Way 
binaries could be inflating the binary fraction of Sextans, which has a lower 
member fraction ($\approx 0.75$) in our selected sample compared to the other 
galaxies.

\begin{figure}[t]
	\includegraphics[height=1.0\hsize,angle=-90]{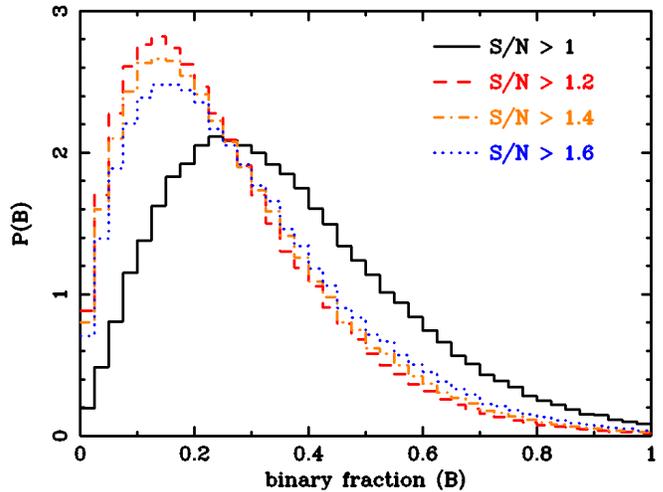}
	\caption{Posterior probability distributions in binary fraction for the 
Carina sample, assuming different S/N cuts. The results are 
essentially the same for S/N cuts greater than 1.2, but the most probable 
inferred binary fraction increases to $\approx$ 0.24 if stars with S/N $> 1$ 
are included (solid curve). This is due to the inclusion of a single star with 
S/N measurements less than 1.2. If this star is a binary, the 
possibility that Carina's low apparent binary fraction is a statistical fluke 
becomes more significant, as binary fractions up to 0.73 are allowed to within 
95\% confidence limits.}
\label{fig:bfposts_sn_carina}
\end{figure}

We can, however, investigate whether the Carina results are sensitive to the 
assumed S/N cutoff. In Figure \ref{fig:bfposts_sn_carina} we plot 
posteriors in the binary fraction of Carina for S/N cutoffs ranging 
from 1 to 1.6. We find that increasing the cutoff from 1.2 does not change the 
inferred binary fraction significantly, although the constraint is weakened 
slightly because the sample size is reduced for higher cutoffs. If the 
S/N cutoff is reduced to 1, however, the most probable inferred 
binary fraction jumps from 0.14 to 0.24. This is due to the inclusion of a 
single star, labeled Car-1543, with a velocity change $\Delta v = 15.7$ 
km~s$^{-1}$ and $\Delta v/\sigma_{2e}$ = 3.6. While it may appear surprising 
that a single star in a sample of $\sim 100$ stars could affect the best-fit 
binary fraction so dramatically, one must keep in mind that only a small 
fraction of the stars exhibit velocity changes beyond the measurement error.  
This accounts for the Poisson-like probability distribution in the binary 
fraction evident in Figure \ref{fig:bfposts_sn_carina}. If Car-1543 is indeed a 
binary, then the inferred binary fraction is less than 0.43 at 68\% CL, and less 
than 0.74 at 95\% CL; this increases somewhat the probability that Carina's 
binary population is consistent with Milky Way field binaries, although the 
best-fit binary fraction is still low.  However, the radial velocity 
measurements for this star, 206.6 km~s$^{-1}$ and 222.3 km~s$^{-1}$ (compared 
to Carina's systemic radial velocity of 223 km~s$^{-1}$), each have relatively 
low S/N equal to 1.09 and 1.19 respectively, and thus may be 
suspect.

A further complication arises in that there are seven stars in the Carina 
sample at magnitudes $m_V \gtrsim 20.3$ (placing them on or near the horizontal 
branch) that show significant velocity variations, greater than 15 km~s$^{-1}$ 
and as high as 40 km~s$^{-1}$.  Only one of these stars lies within the 
color-magnitude cut mentioned above, and this star is a clear non-member in 
view of its low velocity compared to Carina's systemic velocity.  However, 
although none of these stars fall within our sample, such large velocity 
variations at faint magnitudes call for an explanation. None of the stars have 
colors that place them in the RR Lyrae instability strip, so they are unlikely 
to be variable stars.  Instead, four of them lie in the red clump horizontal 
branch region of the color-magnitude diagram, two are clear Milky Way stars due 
to their low velocities, and only one member star lies on the red giant branch.  
However, three stars have S/N less than 1.1, and all seven stars 
have measured velocities with low Tonry-Davis $R$-values, between 4 and 5 
(measurements with $R$-values less than 4 were discarded from the published 
sample to avoid non-Gaussian error due to false-peak selection; see 
\citealt{walker2009}).  Since the velocity variations are quite high compared 
to what might be expected from binaries, we find it likely that these stars 
show velocity variations because of non-Gaussian error related to their 
relatively faint magnitudes.  While the other galaxies in the sample contain 
relatively few stars showing variability at these magnitudes, we note that the 
Carina sample contains 60 multi-epoch stars with $m_V > 20.3$, while the other 
three galaxies combined only have 19 multi-epoch stars in this magnitude range.  
Thus it is perhaps not surprising that Carina has a disproportionately high 
number of stars with large velocity variations at the faint magnitude end of 
the sample.  We consider it unlikely that the velocity variations of stars in 
Carina beyond $m_V > 20.3$ are primarily due to binary motion.

\section{Period distribution constraints}\label{sec:period_distribution}

In the previous section we derived binary fraction constraints by assuming the 
period distribution of each galaxy is identical to that of Milky Way field 
binaries, which takes the form of a log-normal distribution with mean 
$\mu_{\log P} = 2.24$ and dispersion $\sigma_{\log P} = 2.3$. Here we shall 
likewise assume the period distribution to have a log-normal form, but will 
allow $\mu_{\log P}$ and $\sigma_{\log P}$ to vary. Thus the likelihood in each 
star is given by Equation \ref{eq:full_binary_likelihood}, where our set of binary 
parameters $\mathcal{P} = \{\mu_{\log P},\sigma_{\log P}\}$. As in Section 
\ref{sec:binary_fraction_constraints}, we also use magnesium strength 
information to help determine membership and constrain the member fraction $F$.  
Thus our four parameters being varied are the member fraction $F$ and the set 
of binary parameters $\{B, \mu_{\log P}, \sigma_{\log P}\}$.

As in the previous section, we choose a flat prior in the binary fraction $B$ 
over the interval from 0 to 1. Choosing a sensible prior in the period 
distribution parameters $\mu_{\log P}$ and $\sigma_{\log P}$, however, requires 
more careful examination. Binary populations in nearby open clusters have been 
shown to display period distributions that are more peaked compared to that of 
field binaries (\citealt{brandner1998}; \citealt{scally1999}).  This suggests 
that the observed wider period distribution of Milky Way field binaries may 
represent a superposition of narrower distributions resulting from a variety of 
initial star-forming conditions.  However, the period distributions observed in 
open clusters still cover multiple decades of period, and that of classical 
dSphs might be expected to be wider still.  To be conservative, we shall 
consider a uniform prior in the mean log-period $\mu_{\log P}$ over the range 
$\left[0,4\right]$, where the lower limit in this range corresponds to a mean 
period of one year; as we shall see shortly, a binary population with such a 
short mean period would produce velocity variations in excess of what is 
observed in each galaxy, unless the binary fraction is quite small.  We will 
also choose a conservative prior over the period distribution width 
$\sigma_{\log P}$ over the range $\left[1,3\right]$.  The lower limit 
$\sigma_{\log P}=1.0$ is narrower than any open cluster yet observed, and the 
field binary populations in larger dwarf galaxies might be expected to be 
broader than that of clusters; on the other hand, given the relatively small 
size of dwarf galaxies, it would seem unlikely for the period distribution to 
be considerably broader than that of the Milky Way.  In practice, the range of 
the priors will only be relevant when marginalizing over the allowed range of 
$\sigma_{\log P}$ to obtain probability distributions in $B$ and $\mu_{\log 
P}$, as a more conservative prior would produce weaker constraints in these 
parameters.

\begin{figure*}[p!]
	\centering
	\subfigure[Fornax]
	{
		\includegraphics[height=0.38\hsize,width=0.22\hsize,angle=-90]{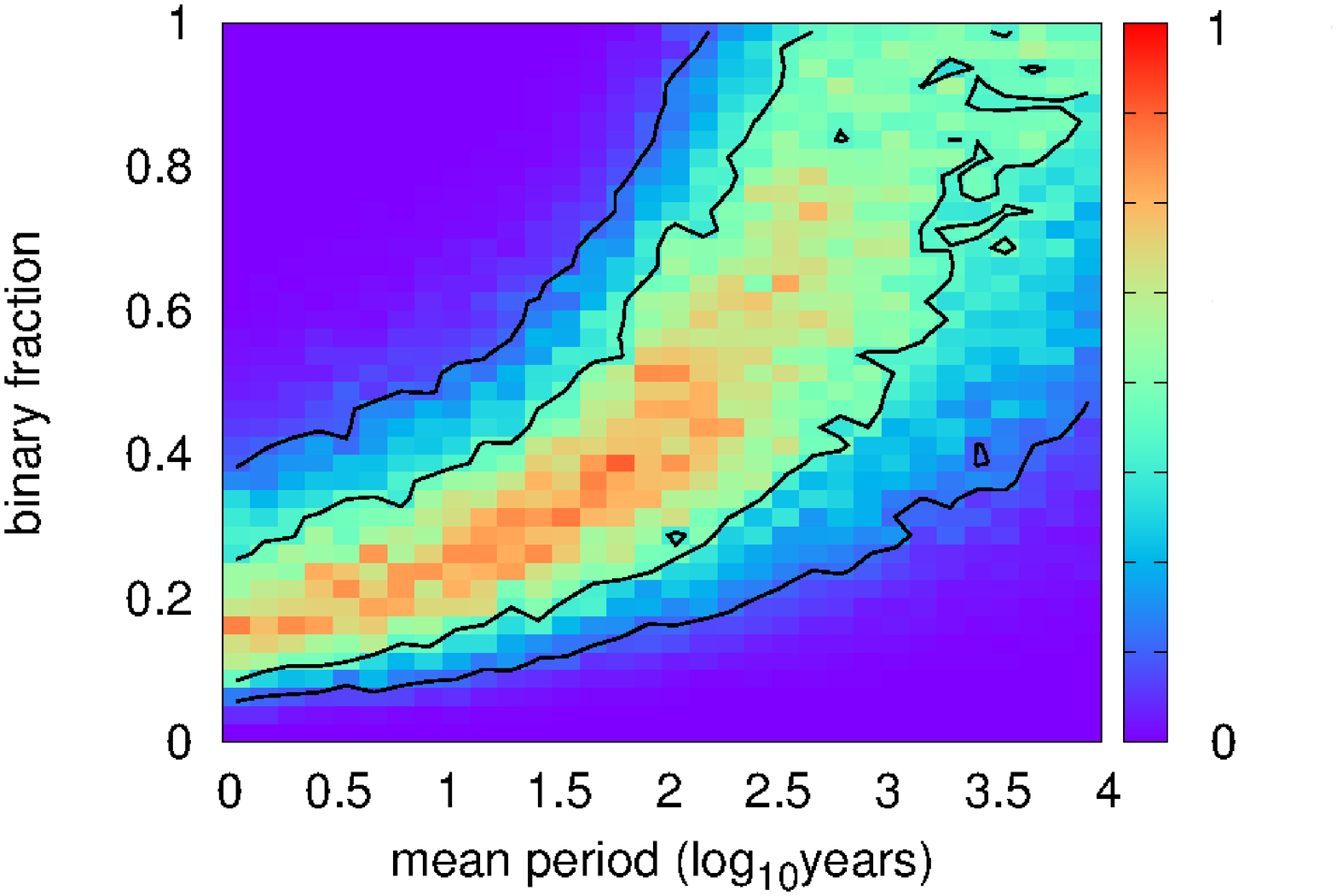}
	}
	\subfigure[Fornax]
	{
		\includegraphics[height=0.38\hsize,width=0.22\hsize,angle=-90]{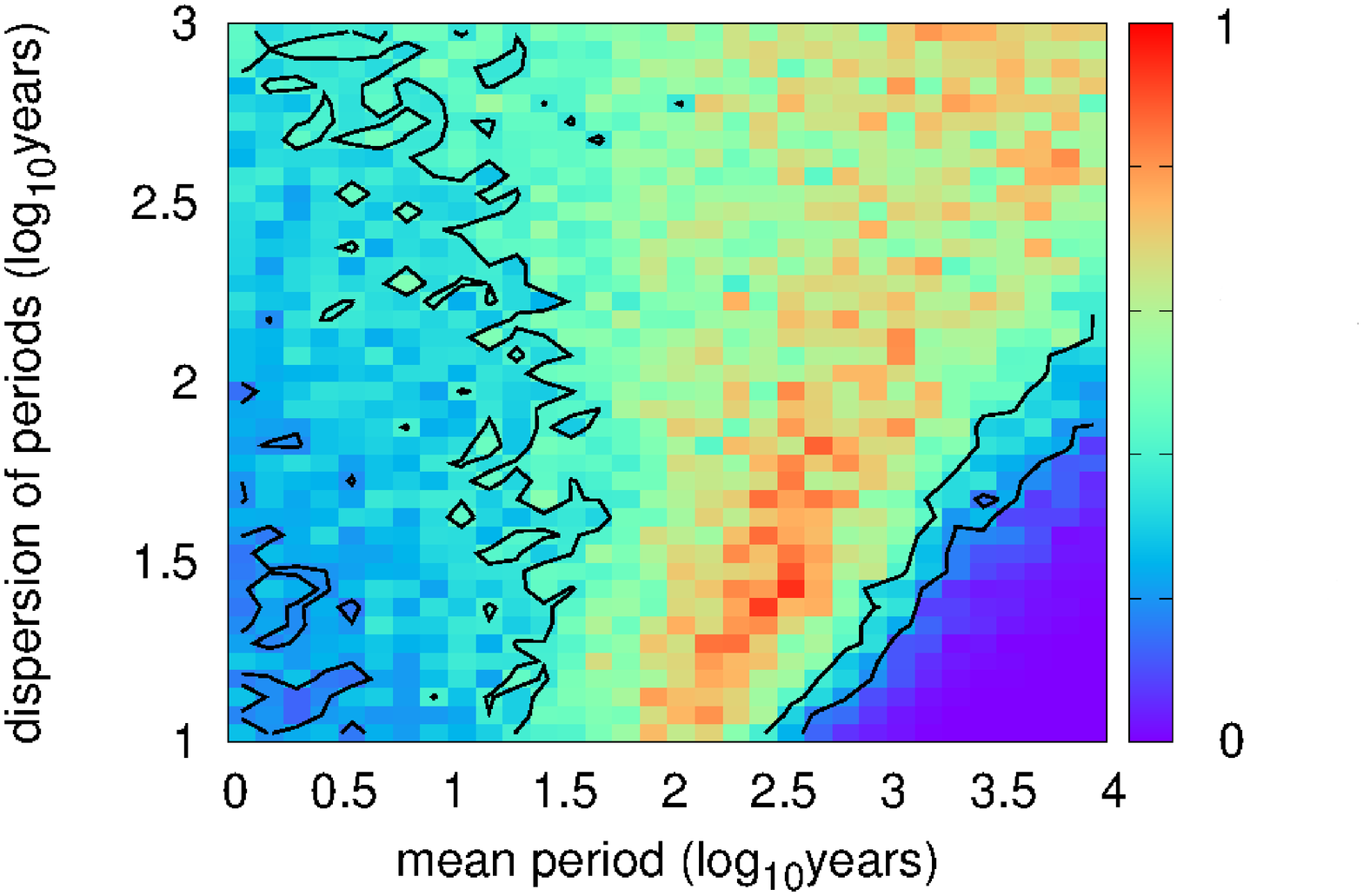}
	}
	\subfigure[Carina]
	{
		\includegraphics[height=0.38\hsize,width=0.22\hsize,angle=-90]{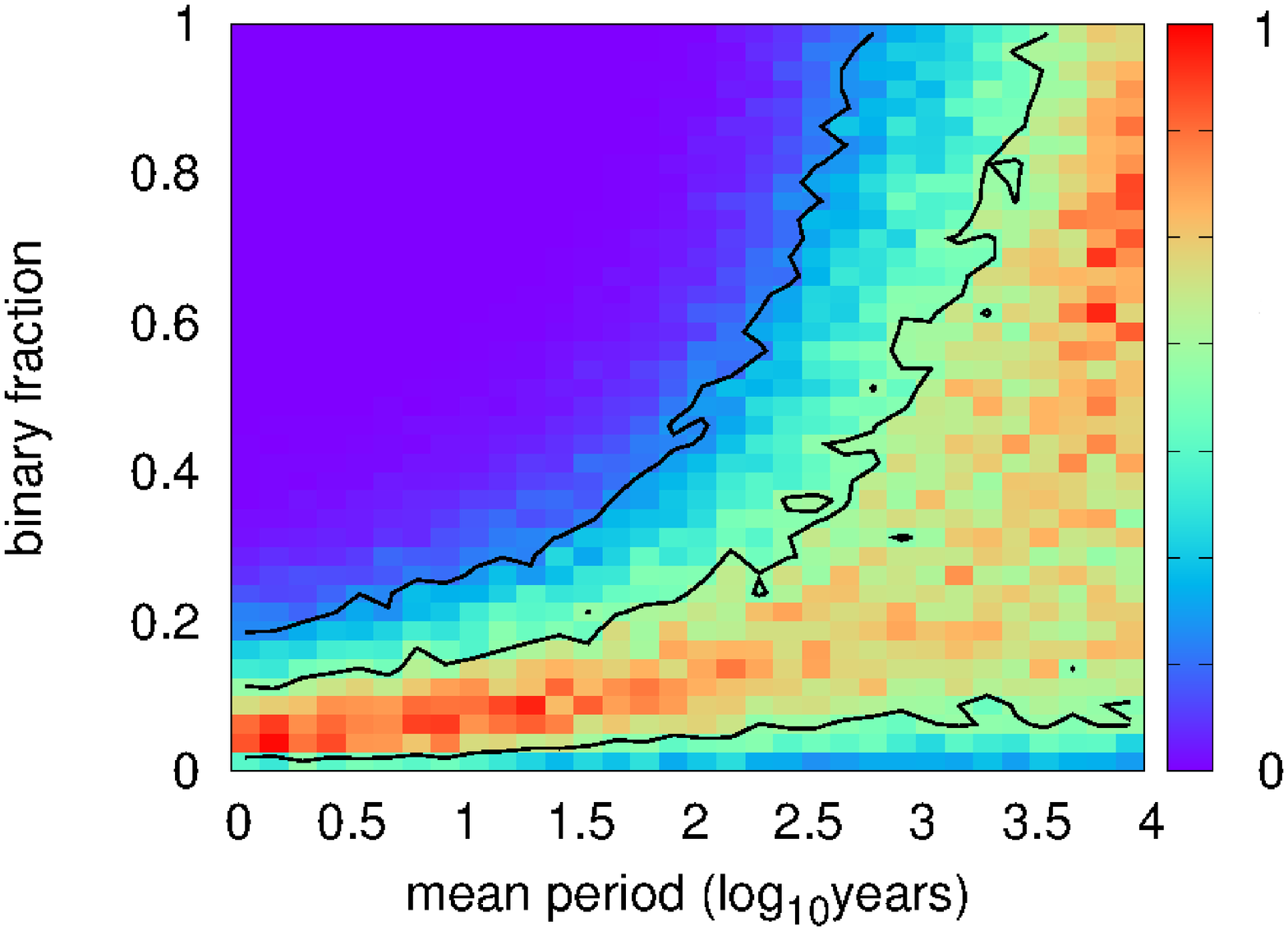}
	}
	\subfigure[Carina]
	{
		\includegraphics[height=0.38\hsize,width=0.22\hsize,angle=-90]{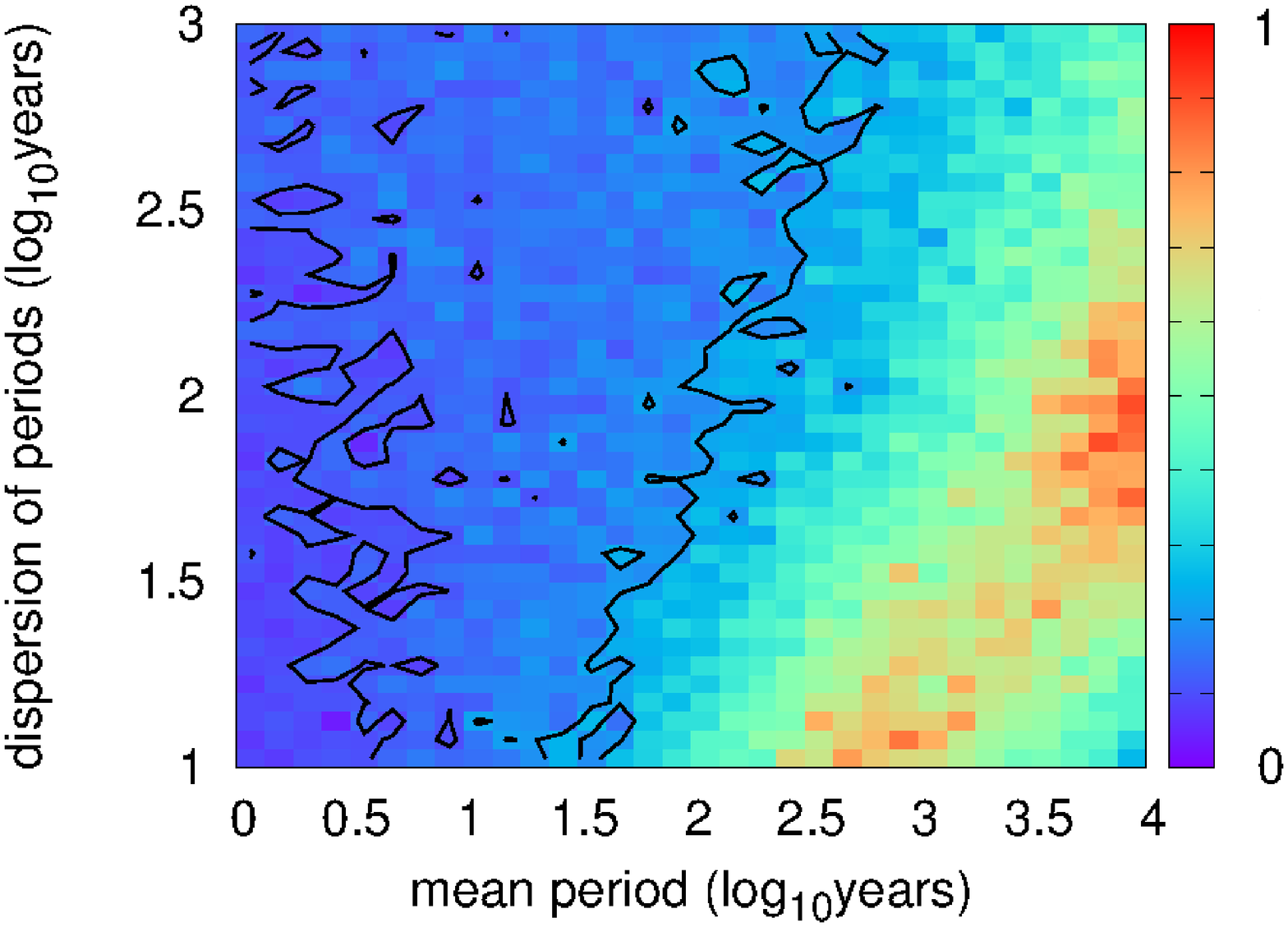}
	}
	\subfigure[Sculptor]
	{
		\includegraphics[height=0.38\hsize,width=0.22\hsize,angle=-90]{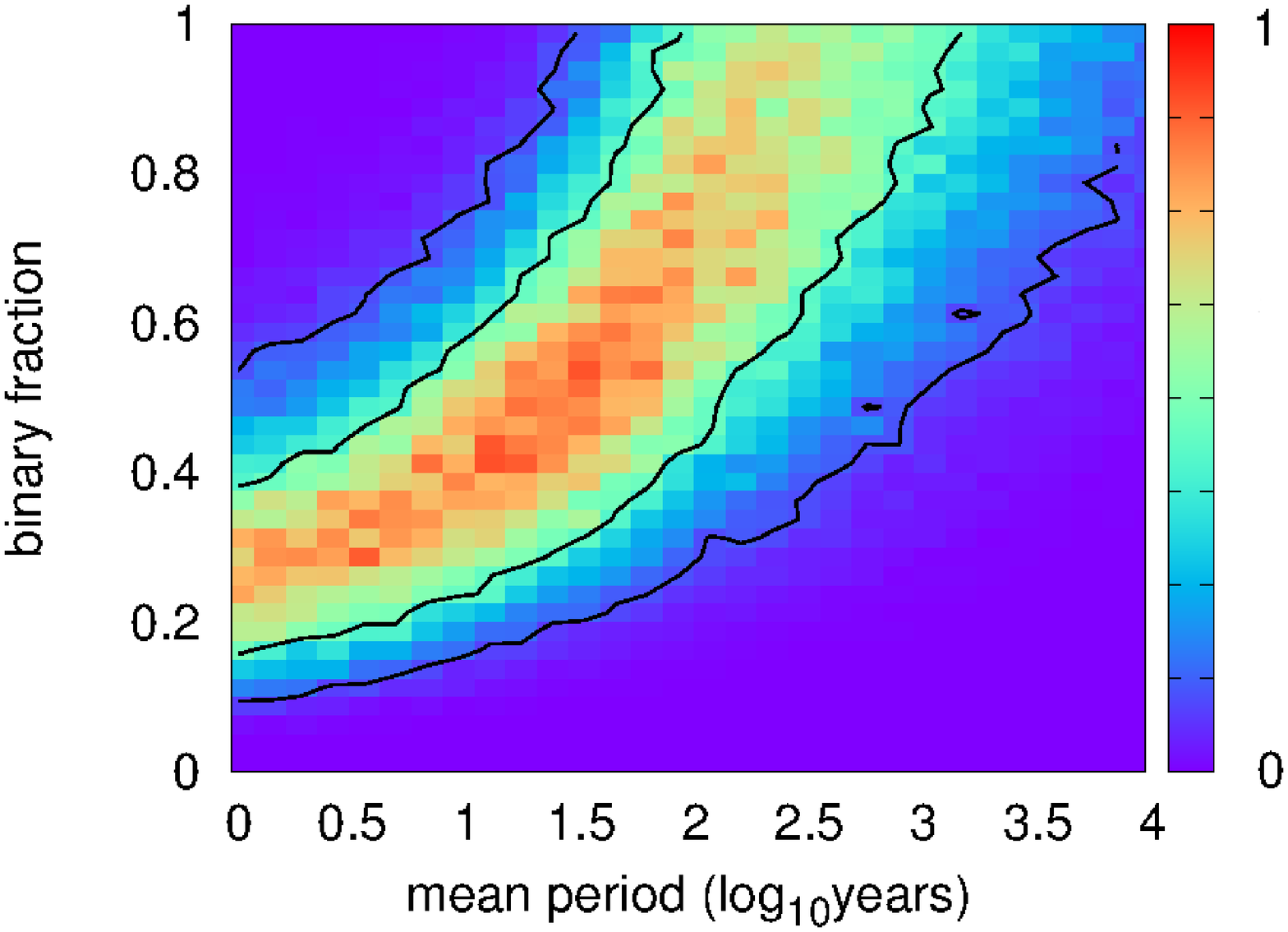}
	}
	\subfigure[Sculptor]
	{
		\includegraphics[height=0.38\hsize,width=0.22\hsize,angle=-90]{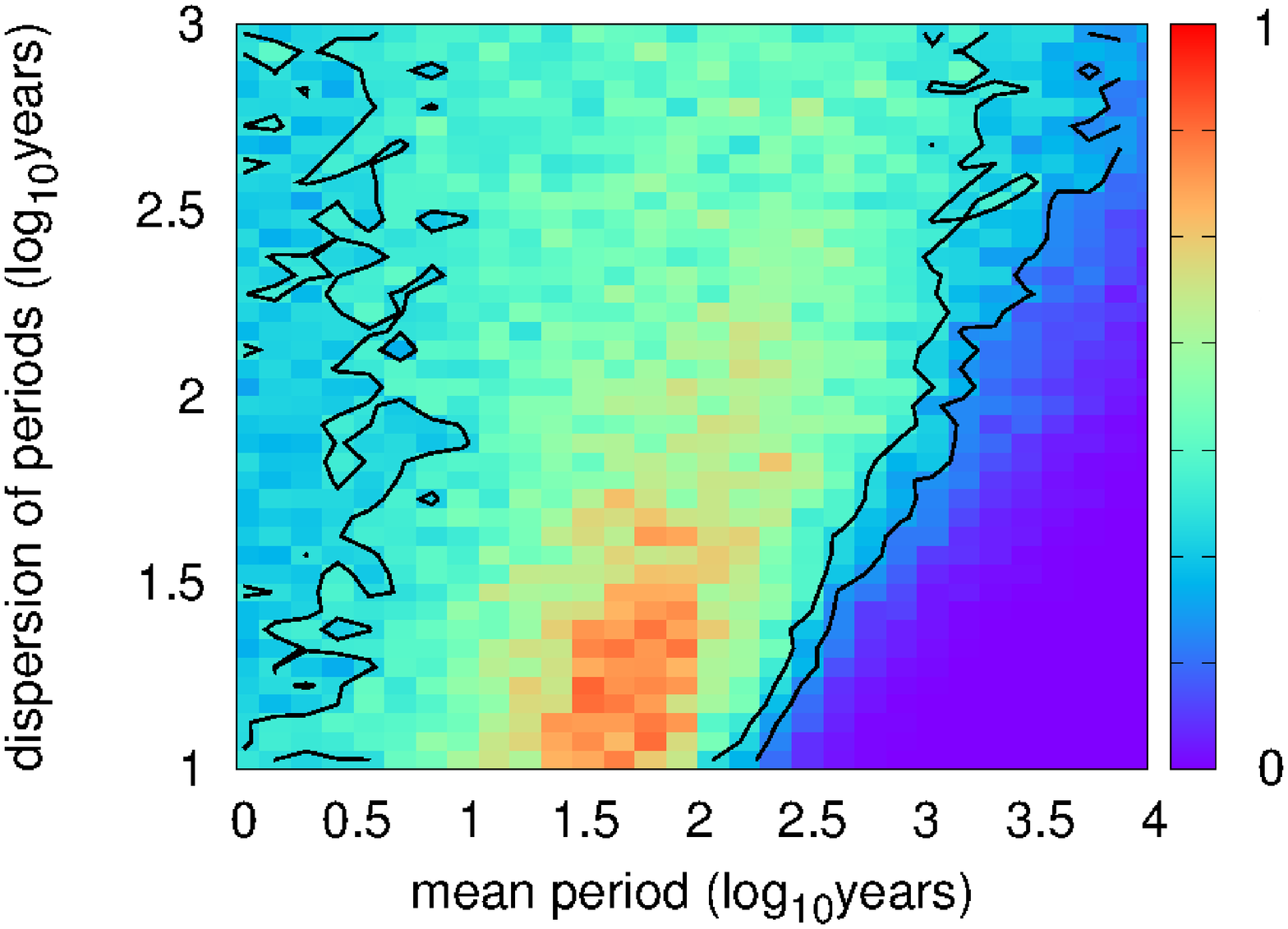}
	}
	\subfigure[Sextans]
	{
		\includegraphics[height=0.38\hsize,width=0.22\hsize,angle=-90]{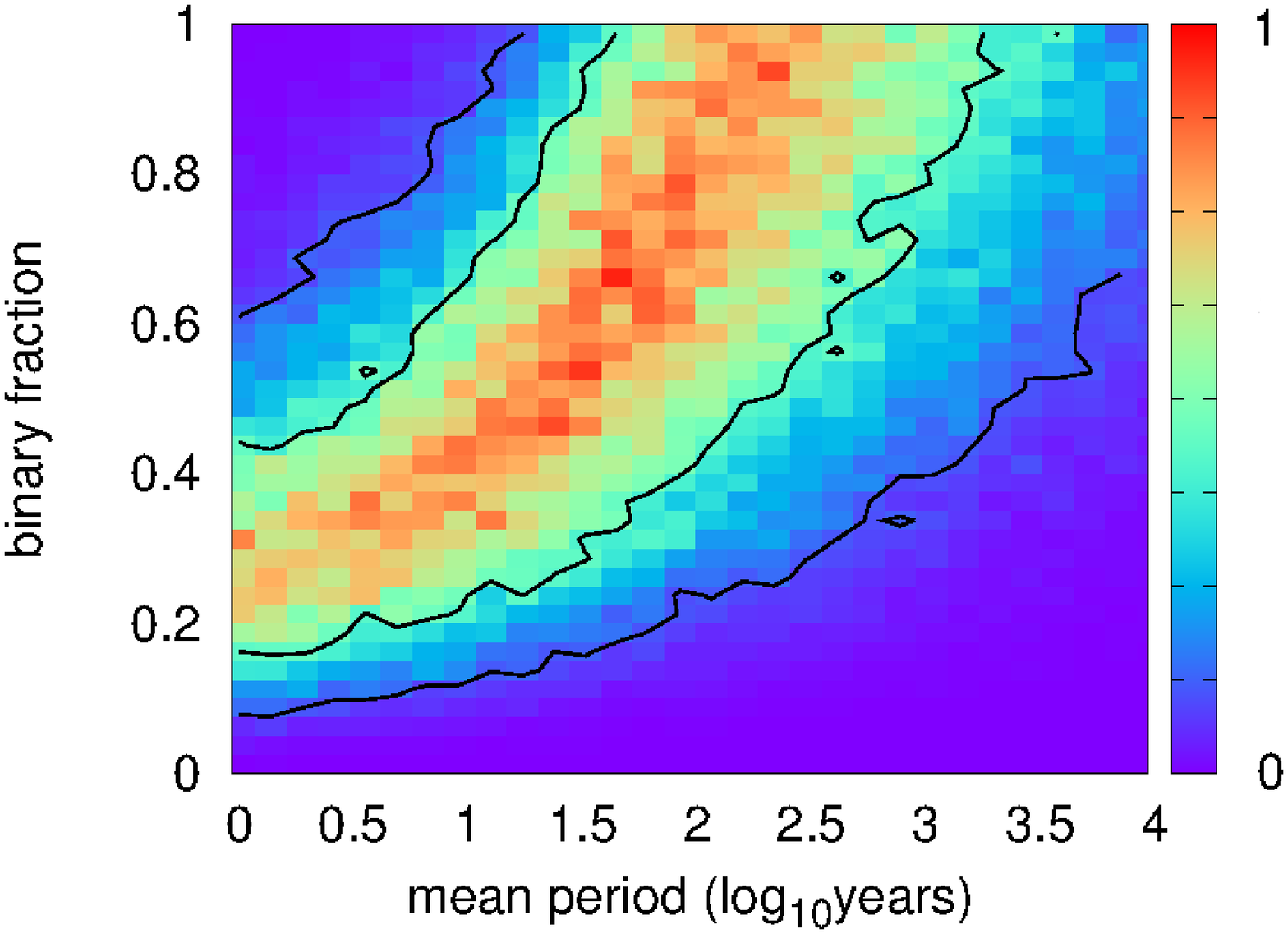}
	}
	\subfigure[Sextans]
	{
		\includegraphics[height=0.38\hsize,width=0.22\hsize,angle=-90]{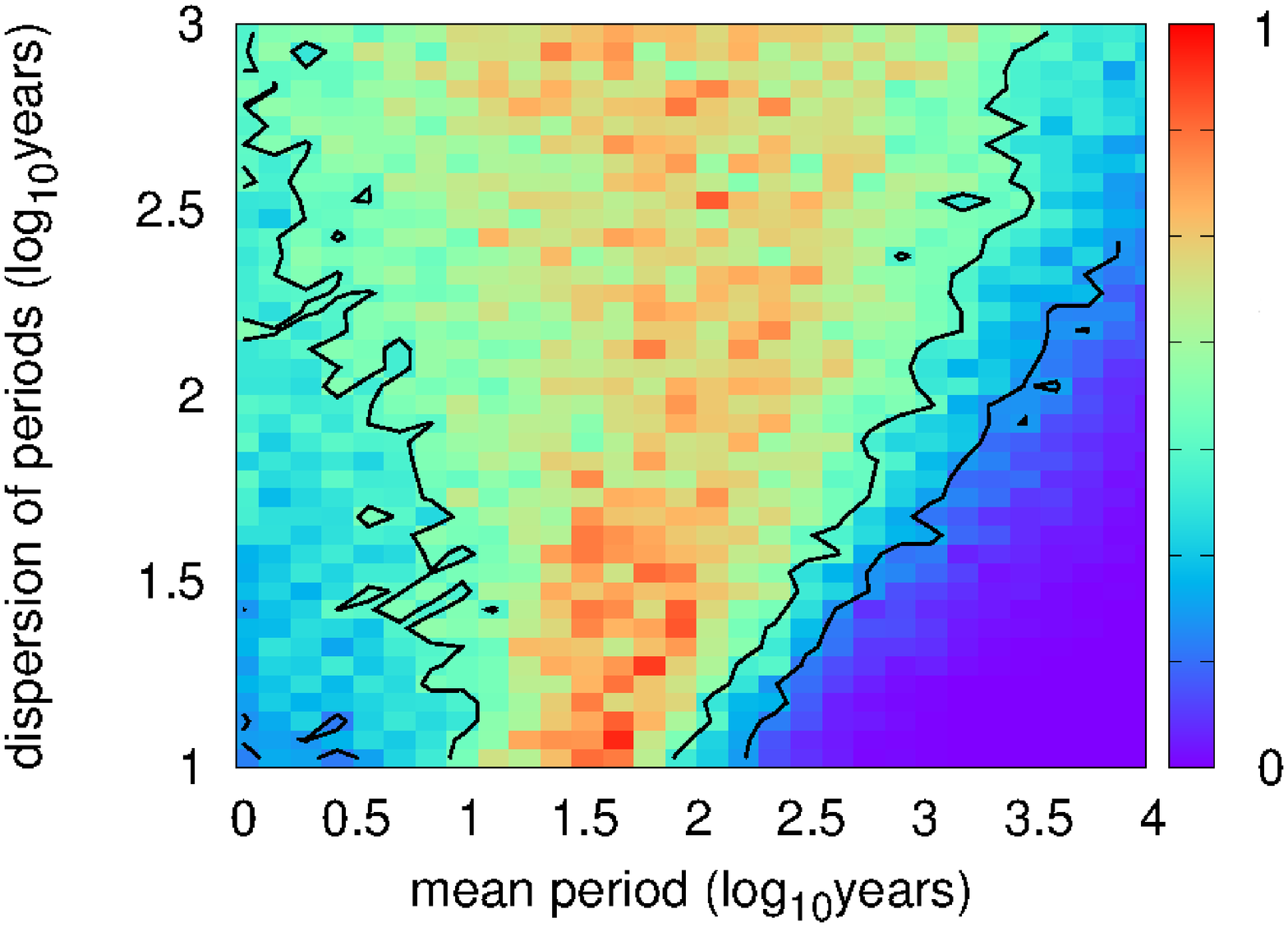}
	}
	\caption{Joint posteriors in binary fraction vs. mean log-period (left) and 
dispersion of log-period vs. mean log-period (right) for each galaxy. A flat 
prior is assumed for each parameter. The inner and outer contours surround the 
regions containing 68\% and 95\% of the total probability, respectively.  Milky 
Way field binaries have a binary fraction $\approx$ 0.5 for solar-type stars, 
with a mean log-period $\mu_{\log P} = 2.24$ and log-dispersion of periods 
$\sigma_{\log P} = 2.3$.  These values lie squarely within the allowed 
parameter space for the Fornax, Sculptor, and Sextans dSphs, but are outliers 
for Carina. Compared to Milky Way binaries, Carina likely has either a smaller 
binary fraction, a longer mean period, or both.}
\label{fig:period_dist_posteriors}
\end{figure*}

\begin{figure*}[t]
	\centering
	\subfigure[2 epochs]
	{
		\includegraphics[height=0.44\hsize,width=0.30\hsize,angle=-90]{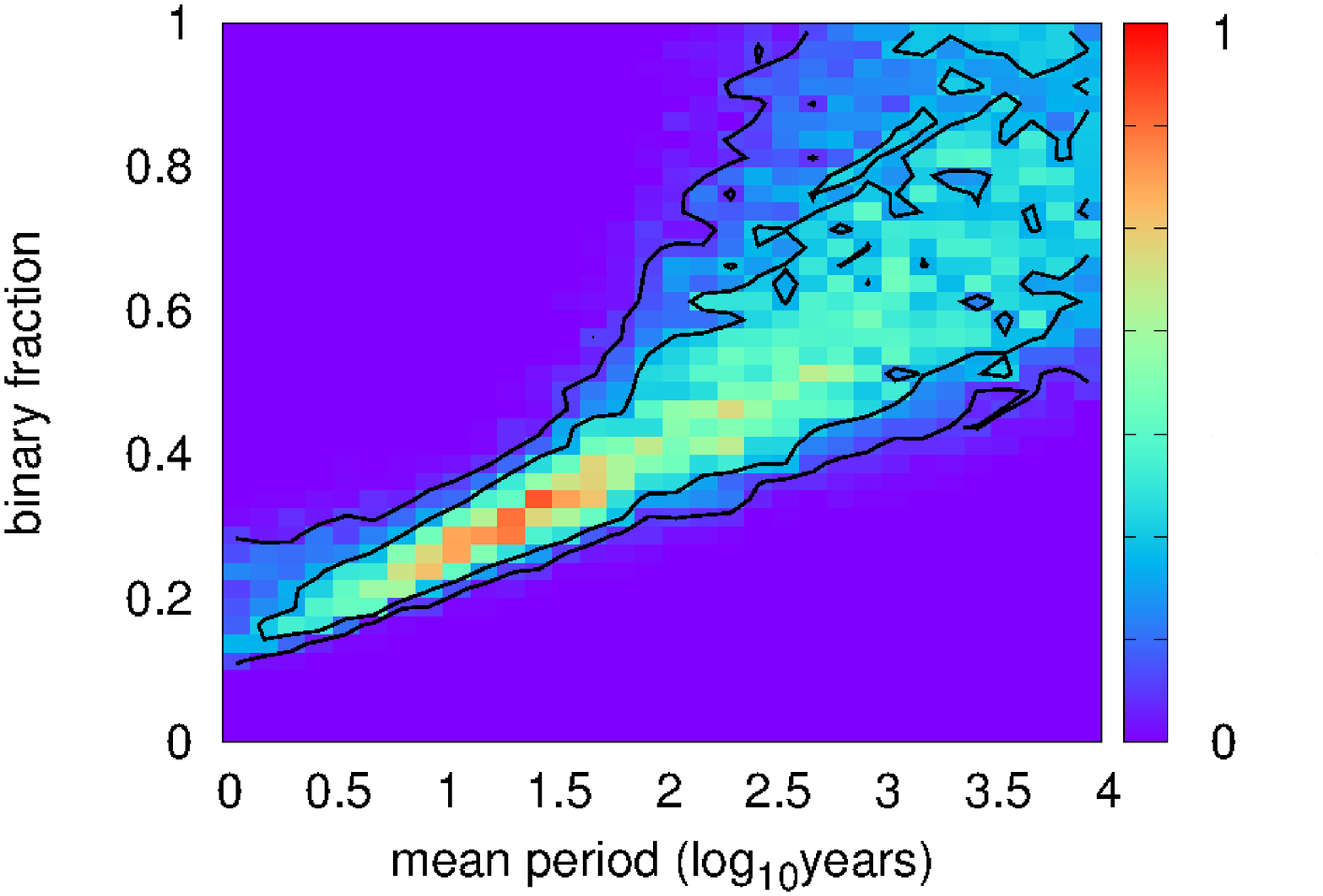}
		\label{fig:sim_b_mup}
	}
	\subfigure[2 epochs]
	{
		\includegraphics[height=0.44\hsize,width=0.31\hsize,angle=-90]{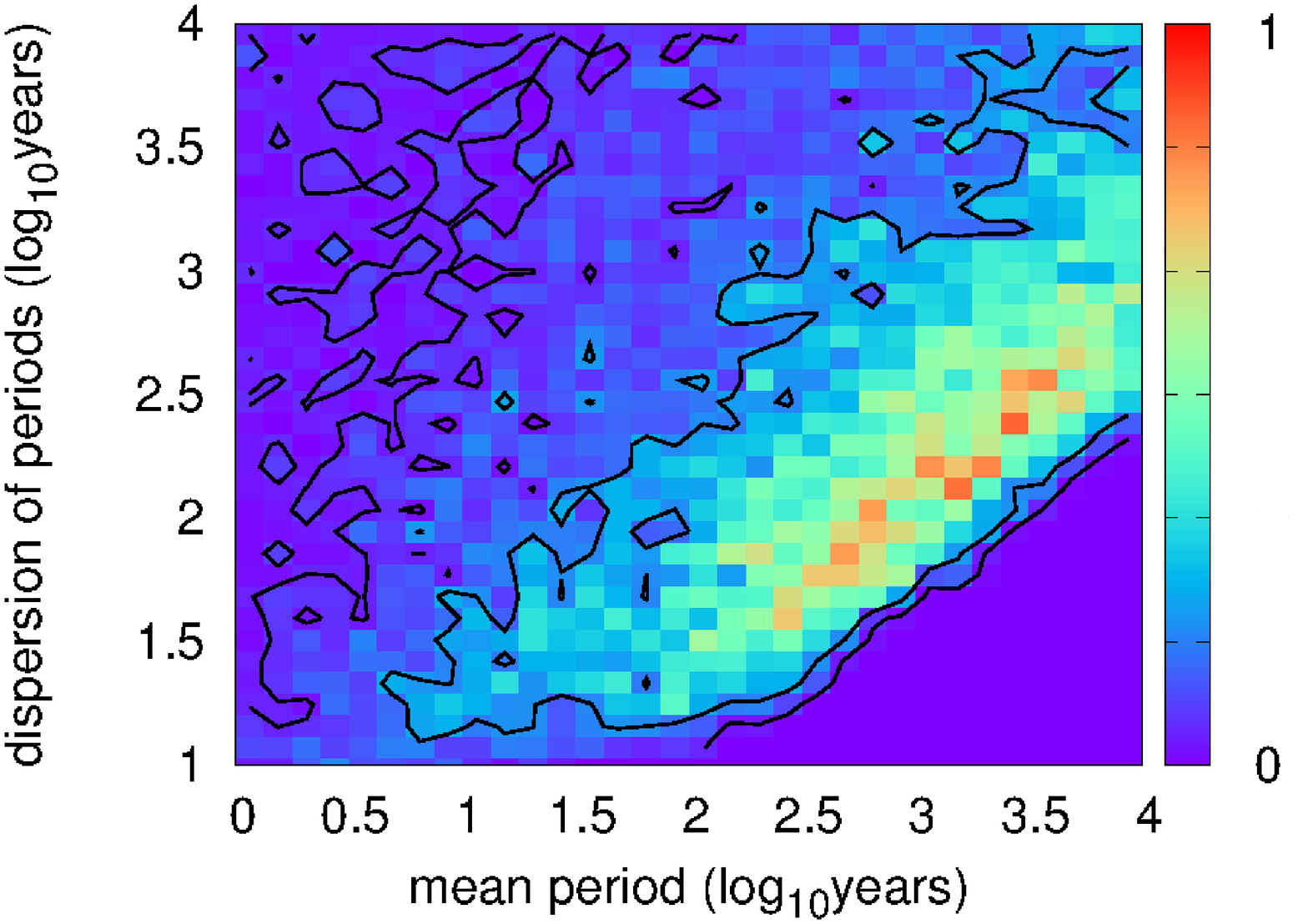}
		\label{fig:sim_mup_sigp}
	}
	\caption{Joint posteriors in mean log-period vs.  binary fraction 
(\ref{fig:sim_b_mup}), and mean log-period vs. dispersion of log-period 
(\ref{fig:sim_mup_sigp}) for a simulated sample of 1500 stars with measurements 
over 2 epochs, and a measurement error of 1.5 km~s$^{-1}$.  The inner and outer 
contours shown here surround the regions containing 68\% and 95\% of the total 
probability, respectively. Even for such a large sample, there is a clear 
degeneracy between the binary fraction and mean log-period, and also between 
the period distribution parameters.}
\label{fig:simulated_period_dist_posteriors}
\end{figure*}

When attempting to constrain the period distribution of a binary population, it 
becomes desirable to include as many repeat measurements as possible for each 
star. Unfortunately however, calculating the binary likelihoods becomes 
computationally expensive for more than two epochs, because the binary tail in 
the likelhood becomes spread out over the velocity space ($\Delta v_1,\Delta 
v_2,\cdots$) and thus a large number of Monte Carlo points are required to 
generate a smooth likelihood, especially for stars with large velocity changes 
in which case the likelihood is quite small.  By running our code in parallel 
over a large number of processors, we can generate likelihoods for stars with 
measurements over as many as four epochs.  Only the Sextans sample contains 
stars with measurements at five epochs or more, albeit a relatively small 
number (seven stars, assuming the epochs to be separated by more than two days).  
For these stars, we discard any epochs occurring after the fourth measurement.

For each galaxy, we calculate the likelihoods for each star in a table of 
values over the period distribution parameters and subsequently interpolate to 
find the binary likelhood for any combination of binary parameters $\mu_{\log 
P}$, and $\sigma_{\log P}$.  After generating the likelihoods for each star, we 
obtain joint posterior probability distributions in $\mu_{\log P}$ vs. $B$, and 
$\mu_{\log P}$ vs.  $\sigma_{\log P}$ by marginalizing over the other 
parameters.  Contour plots of the resulting posteriors for each galaxy are 
shown in Figure \ref{fig:period_dist_posteriors}. For simplicity, each 
distribution is normalized so the highest peak value is equal to 1. For each 
galaxy, we find that the resulting posteriors are degenerate in the three 
binary parameters---for example, a population with a low binary fraction and 
short mean period, and a population with a high binary fraction and long mean 
period, are equally allowed by the data.  These degeneracies result from the 
fact that a large fraction of short-period binaries can be obtained by having a 
large binary fraction, a short mean period (small $\mu_{\log P}$), or a wide 
period distribution (large $\sigma_{\log P}$). The degeneracy between the 
binary fraction and the period distribution parameters has been investigated in 
detail in \cite{minor2010}. In the next section we shall discuss what would be 
required to break this degeneracy and obtain strong independent constraints on 
the binary fraction and period distribution parameters.

By way of comparison, Milky Way field binaries in the solar neighborhood have 
binary parameters $B=0.5$, $\mu_{\log P} = 2.24$, $\sigma_{\log P} = 2.3$.  
This set of parameters lies squarely within the allowed parameter region for 
the Fornax, Sculptor, and Sextans dSphs, indicating that a Milky-Way like 
binary population is compatible with the data in those galaxies. On the other 
hand,the Milky Way parameter set is an outlier for Carina, falling on the edge 
of its allowed parameter space.  Compared to Milky Way field binaries, Carina's 
binary population likely has either a smaller binary fraction ($B < 0.5$), a 
longer mean period ($\mu_{\log P} > 2.3$), or a combination of these.

For each galaxy, we see that parameter regions where the mean period is short 
and the binary fraction is sufficiently high are ruled out, because such binary 
populations would produce greater velocity variations than are observed in the 
data. On the other hand, regions where the mean period is long and binary 
fraction is low are also ruled out, since they would produce fewer velocity 
variations than are observed. Once again the Carina dSph stands out, in that it 
allows small binary fractions even if the mean period is relatively long (up to 
10,000 years). This reflects the lack of velocity variations observed in 
Carina.  However, in the region where the mean period is longer than that of 
the Milky Way ($\mu_{\log P} > 2.24$), the binary fraction is essentially 
unconstrained, for a simple reason: even if the binary fraction is high, binary 
motion can be ``hidden'' if the width of the period distribution is small, 
restricting all binaries to long periods; conversely, if the period 
distribution is broadened, the binary fraction can be reduced so that very few 
short-period binaries would be observed. Thus there is a degeneracy between 
binary fraction and the width of the period distribution that prevents us from 
constraining the binary parameters if the mean period is assumed to be similar 
to the Milky Way or longer. We can say, however, that if the mean period is of 
order 30 years ($\mu_{\log P} \approx 1.5$ or shorter, the binary fraction 
\emph{must} be smaller than 0.4 regardless of the width of the period 
distribution, and is likely smaller than 0.2. We emphasize again that these 
results are somewhat prior-dependent, in that a more conservative prior on 
$\sigma_{\log P}$ would widen the allowed parameter region somewhat.

\section{Binary constraints in simulated data sets}\label{sec:simulations}

\subsection{Degeneracy between binary fraction and period distribution 
parameters}

Since the likelihood method demonstrated in previous sections can constrain the 
parameters characterizing a binary population, it is natural to ask: how large 
a sample, and how many epochs, would be required to obtain strong and 
independent constraints on the binary fraction and period distribution 
parameters? To give some idea of this, we first simulate a binary population 
with a binary fraction $B=0.5$, and period distribution parameters identical to 
that of Milky Way field binaries ($\mu_{\log P}=2.24$, $\sigma_{\log P}=2.3$).  
We assume a measurement error of 1.5 km~s$^{-1}$, which is attainable from 
multi-object spectrographs, especially if measurements are averaged together 
over short timescales. After using a Monte Carlo simulation to generate mock 
stellar line-of-sight velocities in a large sample of 1500 stars, we then 
obtain posterior probability distributions using the same procedure as in 
Section \ref{sec:period_distribution} with identical priors. Posteriors in $B$ 
versus $\mu_{\log P}$ and $\mu_{\log P}$ versus $\sigma_{\log P}$ are plotted in 
Figure \ref{fig:simulated_period_dist_posteriors}.

Figure \ref{fig:sim_b_mup} shows that while the allowed parameter space in $B$ 
and $\mu_{\log P}$ has narrowed considerably compared to the data (Figure 
\ref{fig:period_dist_posteriors}), there is a clear degeneracy between these 
parameters: a high binary fraction and long mean period can produce the same 
observed binary variation, to within the measurement errors, as a low binary 
fraction and short mean period. A similar degeneracy exists between the binary 
fraction and spread of periods $\sigma_{\log P}$. The reason for this 
degeneracy, which is explored in detail in \cite{minor2010}, is that only 
binaries with periods in the short-period tail of the period distribution have 
observable variations beyond the measurement error. Thus the shape of the 
period distribution is not well constrained, as is evident in Figure 
\ref{fig:sim_mup_sigp}, and adjusting the mean period or spread of periods can 
be largely compensated for by changing the binary fraction. We find that this 
degeneracy is still present even if measurements are taken at 3 or 4 epochs, 
indicating that the measurement error is simply too large to constrain the 
shape of the period distribution independently of binary fraction. We note, 
however, that if the mean period of a binary population is shorter than that of 
Milky Way binaries, then the period distribution can be constrained more easily 
since velocity variation is observable for binaries with periods covering a 
larger portion of the period distribution.

Apart from the aforementioned degeneracy in binary fraction and mean period, 
the most probable point in the parameter space of Figure \ref{fig:sim_b_mup} is 
at a low binary fraction and short mean period compared to that of the input 
population. Rather than being largely statistical in nature, this is actually a 
consequence of the prior chosen. We chose a flat prior in the width of the 
period distribution $\sigma_{\log P}$ extending to 3; this allows for periods 
as short as a few days. Binaries with such short periods would exhibit large 
velocity variations, and hence for large period distribution widths, the binary 
fraction would have to be quite small to be in accord with the (simulated) 
data. Thus, by marginalizing over our prior in $\sigma_{\log P}$, we are 
integrating over regions of parameter space where the binary fraction must be 
small, which biases the binary fraction to a low value. The lesson is that one 
must be careful not to choose an overly conservative prior: by choosing a prior 
that allows for too large a period distribution width, one may be inadvertently 
biasing the results.

\begin{figure*}[t]
	\centering
	\subfigure[binary fraction]
	{
		\includegraphics[height=0.31\hsize,width=0.30\hsize,angle=-90]{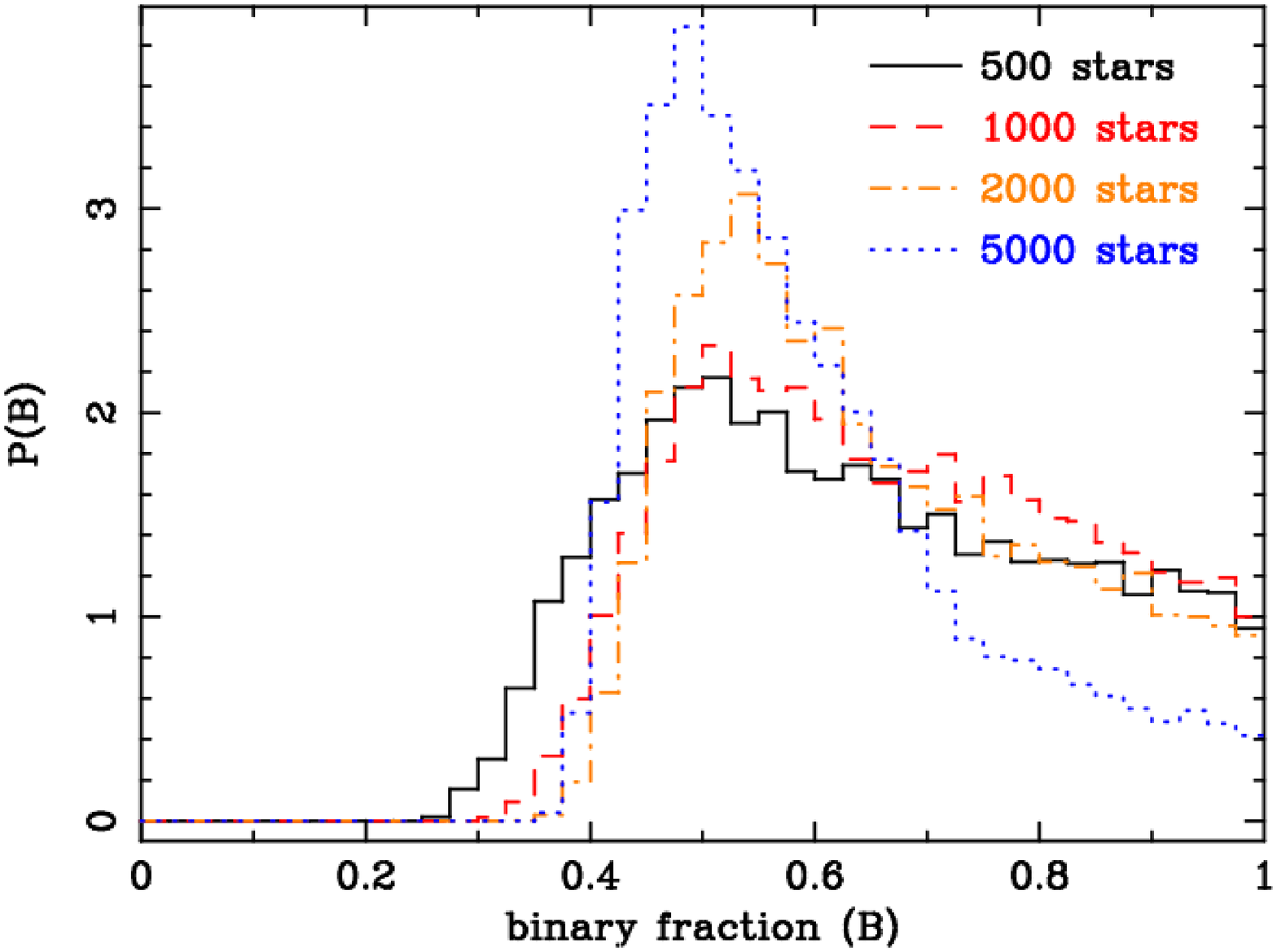}
	}
	\subfigure[mean log-period]
	{
		\includegraphics[height=0.31\hsize,width=0.30\hsize,angle=-90]{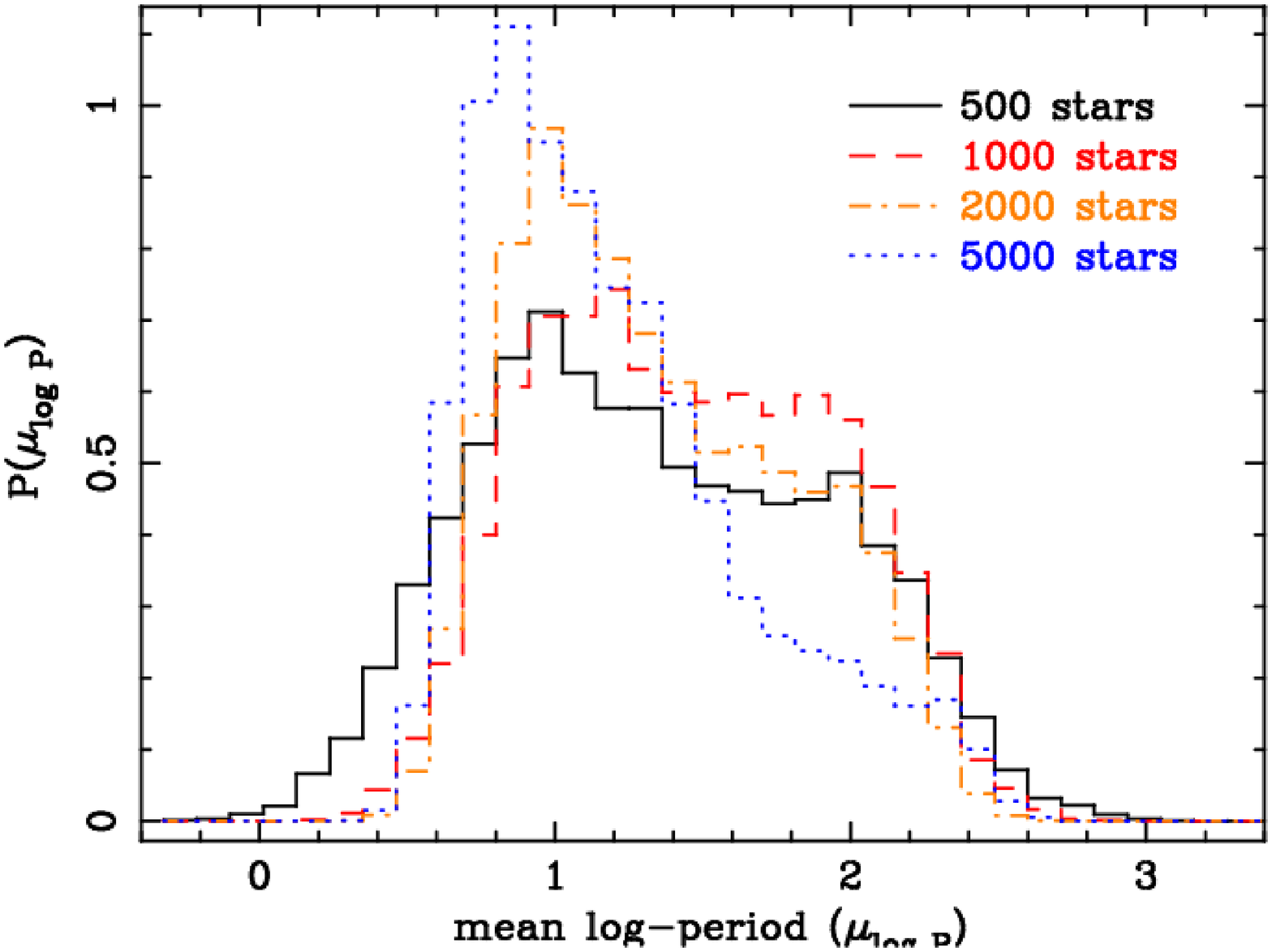}
	}
	\subfigure[log-spread of periods]
	{
		\includegraphics[height=0.31\hsize,width=0.30\hsize,angle=-90]{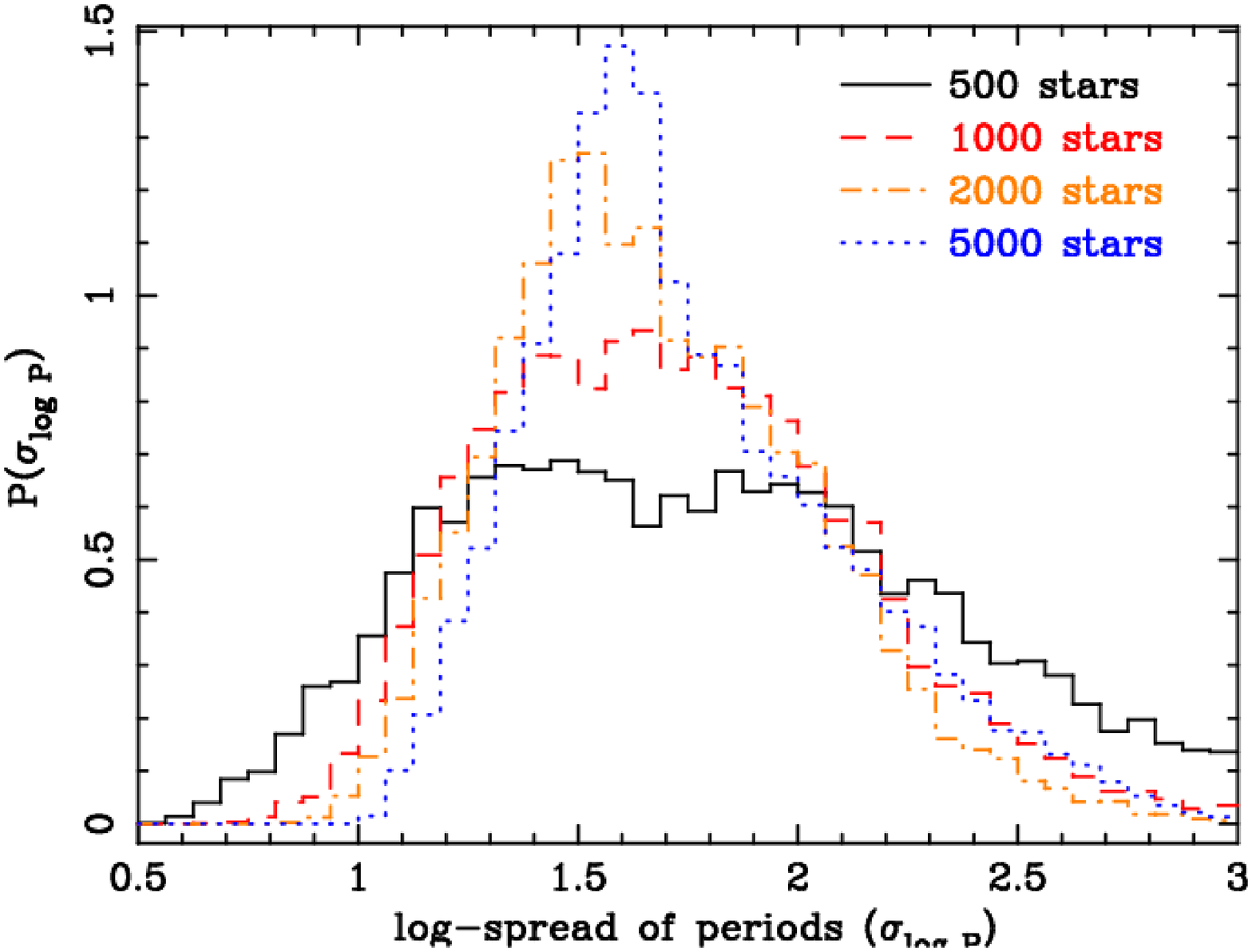}
	}
	\caption{Marginal posterior probability distributions in each binary 
	parameter, for a simulated star population with binary fraction $B=0.5$, 
	mean period of 10 years ($\mu_{\log P}=1$), and log-spread of periods 
	$\sigma_{\log P}=1.5$. Each star has measurements taken at 4 epochs, each 
	spaced 1 year apart, and the measurement error is 0.5 km~s$^{-1}$ for each 
	measurement. As the sample size is increased, the degeracy between the 
	binary parameters becomes weaker and the posteriors peak around the true 
	values in each parameter.}
	\vspace{10pt}
\label{fig:sim4epoch}
\end{figure*}

\subsection{Breaking the degeneracy between binary fraction and period 
distribution parameters in a simulated dataset}\label{sec:breaking_degeneracy}

Before we tackle the question of what would be required to constrain the binary 
fraction and period distribution parameters independently in dSphs, it would be 
reassuring to verify that this is even possible in principle using our 
likelihood approach.  To show this, we simulate a binary population with a mean 
period of 10 years ($\mu_{\log P} = 1$), a spread of periods $\sigma_{\log 
P}=1.5$, and a binary fraction $B=0.5$.  We choose a measurement error of 0.5 
km~s$^{-1}$ and generate simulated samples of 500, 1000, 2000, and 5000 stars with 
this binary population.  Since binary periods are better constrained with a 
large number of epochs, we assume measurements taken at four epochs (the 
largest number allowable at present; computations become prohibitively 
expensive beyond four epochs), each spaced a year apart.

From the simulated line-of-sight velocity data, we obtain posterior probability 
distributions via a nested sampling routine in each parameter $B$, $\mu_{\log 
P}$, $\sigma_{\log P}$, as shown in Figure \ref{fig:sim4epoch}.  Even for a 
sample of 500 stars, the degeneracy between binary fraction $B$ and mean 
log-period $\mu_{\log P}$ is being broken, as evidenced by the peaks appearing 
in the posteriors near their correct values.  The width of the period 
distribution $\sigma_{\log P}$ is more difficult to constrain, and only becomes 
well-constrained beyond 1000-2000 stars. The degeneracy between all three 
parameters is clearly broken beyond 2000 stars.  Note that once the degeneracy 
is broken, the error in the maximum-probability values in each parameter (i.e.  
the difference between the peak value and the ``true'' value) is considerably 
smaller than the width of the posterior distributions, which suggests that the 
Bayesian error may be overly conservative in this case.

\begin{figure}
	\includegraphics[height=1.0\hsize,angle=-90]{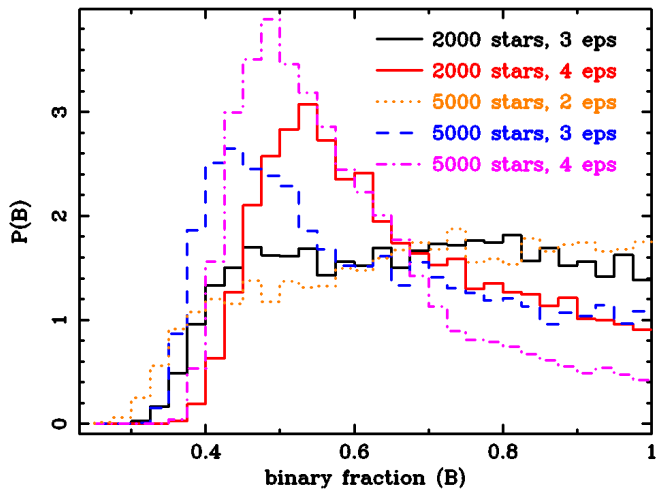}
	\caption{Marginal posterior probability distributions in binary fraction, 
for the same simulated star population as in Figure \ref{fig:sim4epoch} with a 
binary fraction $B=0.5$. We consider samples of 2000 stars with 3- or 4-epoch 
measurements (solid curves) and 5000 stars with 2-, 3-, and 4-epoch 
measurements (dashed, dot-dashed, and dotted curves respectively). Note that 
the binary fraction cannot be recovered with only 2 epochs even in a 5000-star 
sample, since it is still degenerate with the period distribution. In a 
2000-star sample, four epochs are required to constrain the binary fraction 
independently of the period distribution.}
\label{fig:sim234epoch_bf}
\end{figure}

How important is the number of epochs for constraining the period distribution 
parameters? In Figure \ref{fig:sim234epoch_bf} we plot posterior distributions 
in $B$ for 2, 3, and 4-epoch samples of 2000 and 5000 stars. If measurements 
are taken at only two epochs, the degeneracy is not broken even for a 5000-star 
sample. Likewise, if measurements are taken at three epochs, the degeneracy is 
not broken for a 2000-star sample, although it is broken for a sample as large 
as 5000 stars, as evidenced by the peak appearing in the posterior. However, 
the degeneracy is much easier to break in a four-epoch sample; the binary 
fraction is better constrained in a four-epoch sample of 2000 stars compared to 
a three-epoch sample of 5000 stars. We thus conclude that if the measurement 
error is small enough to allow the period distribution to be constrained, 3- or 
4-epoch samples of at least several hundred (and possibly thousands) of stars 
are required to obtain strong and \emph{independent} constraints on the binary 
parameters---in other words, to break the degeneracy between the binary 
fraction and period distribution parameters.

\subsection{Required measurement error for obtaining independent constraints on 
binary parameters}\label{sec:required_error}

For a binary population with a given mean period, what measurement error would 
be required to constrain the period distribution independently of the binary 
fraction? In principle, even if only binaries whose periods lie on the 
short-period tail of the period distribution have observable velocity changes, 
the shape of the period distribution could be inferred if a sufficient number 
of binaries are observed whose periods are well-known. This only holds to the 
extent that the assumption of a log-normal period distribution is accurate, so 
that the shape of the period distribution can be extrapolated to longer 
periods.  Unfortunately however, with only 2-4 epochs, binary periods of 
individual stars cannot be determined with great accuracy.  Furthermore, the 
likelihood approach is not computationally feasible at present with more than 4 
epochs, because of the enormous number of Monte Carlo points required to 
generate smooth likelihoods.  It is thus reasonable to expect that to constrain 
the period distribution well using the likelihood approach, observing only 
binaries on the short-period tail of the period distribution is not sufficient.  
We make the assumption that binary periods nearly as long as the mean period 
($\approx$ 180 years for Milky Way field binaries) should be observable, i.e.  
should have observable velocity variations, in order to constrain the period 
distribution parameters well.  We can get an approximate idea of the 
measurement error required for this by considering typical velocity amplitudes 
associated with binaries with a given orbital period. In \cite{minor2010}, a 
formula is derived for the velocity amplitude of a binary as a function of 
period, which is averaged over the possible orientations, mass ratios, and 
eccentricities of binary systems (assuming Milky-Way like PDFs in these 
parameters):

\begin{equation}
\label{envelope_estimate}
v_{max} \approx (5.7 ~ \textrm{km~s$^{-1}$}) 
\left(\frac{M/M_\odot}{P/\textrm{year}}\right)^{\frac{1}{3}}
\end{equation}
where $M$ is the mass of the primary star. If we assume $M \approx 0.8 M_\odot$ which 
is typical of old stellar populations in dwarf spheroidals, we find that 
binaries with periods longer than a few decades will yield velocity amplitudes 
less than 2 km~s$^{-1}$, and thus would be difficult to distinguish from a measurement 
error of 1-2 km~s$^{-1}$. Binaries with a period of 180 years would have velocity 
amplitudes of $\approx$ 0.9 km~s$^{-1}$ on average. However, there is a further 
difficulty: if measurements are taken over the course of a few years, binaries 
of such long periods would traverse only a small fraction of their orbits and 
thus would exhibit velocity changes much smaller than their overall velocity 
amplitude. As a rough approximation, if we assume a circular orbit, it can be 
shown that the rms change in line-of-sight velocity after a time $\Delta t$ is 
given by the simple equation

\begin{equation}
\Delta v_{rms} = v_{max} \left|\sin \frac{2\pi \Delta t}{P}\right|,
\label{eq:v_rms}
\end{equation}
where we have averaged over all possible orbital angles. In an approximate 
sense, velocity changes will be observable if they are larger than the 
measurement error between two pairs of velocity measurements. If the 
measurement error $\sigma_m$ is roughly the same for either measurement, then 
the error in the velocity difference $\Delta v$ is $\sigma_{2e} = 
\sqrt{2}\sigma_m$. Using this together with eqs. \ref{envelope_estimate} and 
\ref{eq:v_rms}, we find the measurement error must be smaller than

\begin{equation}
\sigma_m \approx 4.0~km~s^{-1}\left(\frac{M}{P}\right)^{\frac{1}{3}}\left|\sin\frac{2\pi\Delta t}{P}\right|.
\label{eq:sigm_vs_logp}
\end{equation}

\begin{figure}
	\includegraphics[height=1.0\hsize,angle=-90]{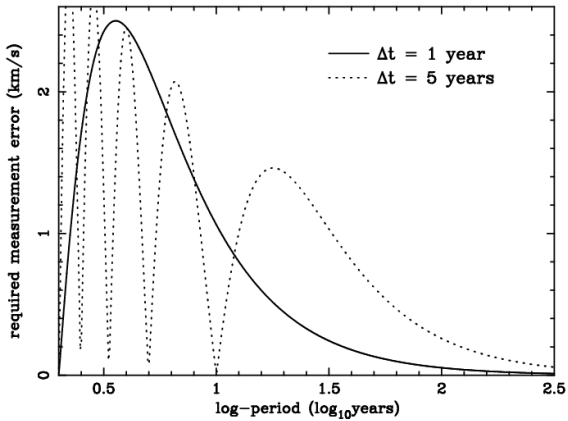}
	\caption{Approximate measurement error required to observe velocity variation in binaries of a given orbital period, over a timescale of 1 year (solid curve) and 5 years (dotted curve). As can be seen from the dotted curve, to observe orbital periods comparable to the mean period of Milky Way field binaries ($\log_{10}P = 2.24$), a measurement error on the order of 0.1 km~s$^{-1}$ would be required over a timescale of several years.}
\label{fig:sig_vs_logp}
\end{figure}

This is the maximum (approximate) measurement error with which one can observe 
binary motion for a binary with period $P$ over a timescale $\Delta t$.  In 
Figure \ref{fig:sig_vs_logp} we plot the required measurement error as a 
function of log of period over timescales of 1 year and 5 years. Note that the 
required measurement error is very small near periods for which the timescale 
$\Delta t$ is an integer multiple of one-half the period, since the velocity 
change is typically small for these timescales. However if the period is long 
enough, the required measurement error decreases monotonically. For binaries 
with periods equal to the Milky Way's mean period ($\log P = 2.24$), however, 
the maximum measurement error for observing binary motion over 5 years is 
$\sim$0.1 km~s$^{-1}$, and much smaller over a timescale of 1 year. We can therefore 
expect that if a binary population has a mean period similar to that of Milky 
Way field binaries, it would be extremely difficult to constrain its period 
distribution parameters independently of binary fraction if the measurement 
error is significantly larger than 0.1 km~s$^{-1}$, unless the measurements are taken 
over timescales considerably longer than 5 years. The prospects are improved if 
the mean period is shorter; for example if the mean period is 10 years, Figure 
\ref{fig:sig_vs_logp} suggests that the period distribution can be constrained 
independently of binary fraction if the measurement error is less than roughly 
1 km~s$^{-1}$.

It should be emphasized that Equation \ref{eq:sigm_vs_logp} gives only an 
approximate idea of the measurement error required to break the degeneracy 
between binary fraction and the period distribution parameters; it says nothing 
about the number of stars and epochs that would be required to break this 
degeneracy, even if the measurement error is small enough. As we explored in 
Section \ref{sec:breaking_degeneracy} for the case of a relatively short mean 
period, at least three or four epochs and several hundred up to several 
thousands of stars would be required, but this requirement would likely vary 
depending on the actual period distribution and measurement errors in question.

Finally, we note that the task of constraining the binary period distribution 
would be considerably easier if the binary fraction were known \emph{a priori}.  
This is possible in globular clusters, where the binary fraction can be 
constrained by photometry (\citealt{sollima2007}, \citealt{sollima2012}).  
Because the stars in the cluster have nearly identical ages and metallicities, 
the main sequence is well-defined and binaries will lie off the main-sequence 
for single stars due to the combined light of the primary and secondary star.  
After the binary fraction is constrained from photometry, this constraint could 
be included as a prior in a likelihood analysis following a radial velocity 
variability survey to constrain the period distribution in the cluster.  
Unfortunately in dwarf spheroidal galaxies, constraining the binary fraction 
through photometry is a much more difficult task due to the complicated 
distribution of stellar ages and metallicities in each galaxy.

\section{Discussion}\label{sec:discussion}

We demonstrated in Section \ref{sec:binary_fraction_constraints} that the 
Carina dSph sample is nearly devoid of short-period binaries, with a binary 
fraction less than 0.5 to within 90\% confidence limits if a Milky Way-like 
period distribution is assumed.  While the reason for Carina's apparent 
shortage of close binaries is unclear, several factors suggest themselves.  
Photometric studies show Carina to have a bursty star formation history 
(\citealt{smecker-hane1996}; \citealt{hurley-keller1998}), with two predominant 
star populations: approximately 25\% of its stars are older than 10 Gyr, while 
the remaining majority are intermediate age stars with ages in the range of 4-7 
Gyr. The question is therefore, what initial conditions during its relatively 
short period of star formation might have prevented close binaries from 
forming?  Potential factors are the initial temperature of the star-forming 
clouds prior to collapse (\citealt{sterzik2003}), as well as the presence of 
magnetic fields and strong radiative feedback, which can all play a role in 
inhibiting binary star formation (\citealt{price2010}).  While Carina's low 
metallicity compared to the Milky Way might conceivably affect radiative 
feedback and hence suppress binary star formation, we note that the Sculptor 
and Sextans dSphs also have relatively low mean metallicities, and their 
best-fit binary fractions are both higher than 0.5. Thus, its low metallicity 
alone is unlikely to explain the apparent lack of binaries in the Carina dSph.

It is worth noting that a recent spectroscopic study of metal abundances in 
Carina (\citealt{lemasle2012}) showed little evidence of iron enrichment from 
type Ia supernovae during its intermediate period of star formation, even 
though a few Gyr should have been a long enough timescale for SNe Ia to 
contribute iron to the interstellar medium while star formation was still 
taking place.  While there is still some uncertainty in the timescale for star 
formation of Carina's intermediate age population, this lack of evidence for 
SNe Ia would be expected if Carina is indeed deficient in short-period 
binaries, as the present study suggests.

As we cautioned in Section \ref{sec:carina_low_bf}, however, the possibility 
that Carina's apparently low fraction of binaries may be a statistical 
coincidence increases somewhat if the star Car-1543 is a binary, although the 
best-fit binary fraction is still low.  However, this star's observed velocity 
variation may be explained by low signal-to-noise measurements (S/N $\sim$1.1).  
Furthermore, stars near the horizontal branch region also exhibit velocity 
variations, although most or all of these are probably explained by poor 
spectrum cross-correlation fits related to their faint magnitudes.  

\section{Conclusions}\label{sec:conclusion}

We have developed a general methodology for constraining properties of binary 
star populations from multi-epoch line-of-sight velocity measurements in dwarf 
galaxies. This method has been applied to data from the Magellan/MMFS sample of 
\cite{walker2009} in the Carina, Fornax, Sculptor, and Sextans dwarf spheroidal 
galaxies to find constraints in their binary fraction and period distribution 
parameters.  To obtain the best possible binary constraints, we have also 
re-derived the measurement errors in the sample by extending the error model of 
\cite{walker2007} to account for binary orbital motion. The best-fit binary 
fractions in each galaxy, assuming a Milky Way-like period distribution, are 
listed in Table \ref{tab:binary_fraction}, with probability distributions in 
binary fraction plotted in Figure \ref{fig:bfposts_alldwarfs_sn1.2}; more 
general probability distributions in the period distribution parameters and 
binary fraction are plotted in Figure \ref{fig:period_dist_posteriors}. We 
conclude with the following points:

1. If a Milky Way-like period distribution is assumed in each galaxy, the 
Fornax, Sculptor, and Sextans dSph galaxies have binary fractions that are 
roughly consistent with that of Milky Way field binaries, whereas the Carina 
dSph is apparently deficient in binaries compared to the Milky Way field.  
Carina's inferred binary fraction is less than 0.5 at the 90\% confidence 
level, with a best fit value of $0.14^{+0.28}_{-0.05}$; thus a Milky Way-like 
binary fraction of $\approx 0.5$ in Carina is statistically unlikely. Relaxing 
the assumption of a Milky Way-like period distribution, the lack of observed 
binary velocity variation in Carina could be explained by either a small binary 
fraction, or a long mean period, compared to the other galaxies in the sample 
(see Figure \ref{fig:period_dist_posteriors}).

2. With the exception of Carina, the published measurement errors in the 
Magellan/MMFS sample of \cite{walker2009} underestimate the Gaussian 
measurement error somewhat, as can be seen in Table \ref{tab:median_errors}.  
This is most striking in the case of the Fornax dSph, where the median 
measurement error is larger than the published value by a factor of $\approx$ 
55\%. While this difference of $\approx$ 0.6 km~s$^{-1}$ in the median error is small 
compared to Fornax's velocity dispersion of $\approx$ 12 km~s$^{-1}$, the extra 
measurement error should be taken into consideration when applying mass models 
to Fornax.  This may be particularly of concern if higher moments in the 
velocity distribution are used, since stars lying on the tail of Fornax's 
velocity distribution may have unaccounted-for measurement error if the 
published errors are used.

3. While multi-epoch surveys can in principle produce independent constraints 
on a galaxy's binary fraction and the period distribution parameters, these 
parameters are unfortunately degenerate given the present measurement errors.  
For example, a binary population with either a low binary fraction and short 
mean period, or a high binary fraction and long mean period, can both produce 
the same velocity variations observed in the data equally well. If a galaxy's 
mean binary period is roughly comparable to that of the Milky Way, strong 
\emph{independent} constraints on the binary parameters can only be obtained 
with a measurement error of order $\sim$0.1 km~s$^{-1}$ or smaller, obtainable only by 
a high-resolution spectrograph. In the near future, the best possible binary 
constraints may be obtained in globular clusters where the binary fraction can 
be estimated independently through photometry, followed by a multi-epoch radial 
velocity survey to constrain the binary period distribution.

4. Although the degeneracy between the binary parameters cannot be broken with 
measurement errors presently obtainable by multi-object spectrographs,
meaningful comparisons between different binary populations can still be made.  
This was demonstrated in Figure \ref{fig:period_dist_posteriors}: although 
independent constraints on the binary fraction and mean period cannot be 
obtained, it is evident that the Carina dSph occupies a different region of 
parameter space compared to the other galaxies in the sample. Larger 
multi-epoch surveys in dwarf galaxies will narrow the parameter space further 
to allow even more robust comparisons between galaxies. By comparing the 
allowed parameter space for different galaxies, we can ultimately address the 
question of whether field binary populations are universal in nature, or vary 
depending on the initial conditions present during the galaxy's epoch of star 
formation.  Large multi-epoch surveys, in concert with statistical methods like 
that demonstrated here, will provide a powerful test of star formation theories 
in the future.

\section*{Acknowledgements}
The author would like to thank Matt Walker for generously providing the error 
model data without which this work would not have been possible. I would also 
like to thank Erik Tollerud for providing useful comments on the manuscript, 
and Manoj Kaplinghat, James Bullock, Erik Tollerud, and Greg Martinez for 
providing valuable feedback and support throughout the course of this project.

This research was supported, in part, by a grant of computer time from the City 
University of New York High Performance Computing Center under NSF Grants 
CNS-0855217, CNS-0958379 and ACI-1126113.

This work was supported in part by NSF grant AST-1153335.

~ ~ ~ ~ ~ ~ ~ ~ ~ ~ ~ ~ ~ ~ ~ ~ ~ ~ ~ ~ ~ ~ ~ ~ ~ ~ ~ ~ ~ ~ ~ ~ ~ ~ ~ ~ ~ ~ ~

\bibliographystyle{apj}

\begin{thebibliography}{44}
\expandafter\ifx\csname natexlab\endcsname\relax\def\natexlab#1{#1}\fi

\bibitem[{{Bate}(2009)}]{bate2009} {Bate}, M.~R. 2009, \mnras, 392, 590
\bibitem[{{Bate}(2012)}]{bate2012} ---. 2012, \mnras, 419, 3115
\bibitem[{{Brandner} \& {Koehler}(1998)}]{brandner1998} {Brandner}, W. \& {Koehler}, R. 1998, \apjl, 499, L79+
\bibitem[{{de Boer} {et~al.}(2012{\natexlab{a}}){de Boer}, {Tolstoy}, {Hill}, {Saha}, {Olsen}, {Starkenburg}, {Lemasle}, {Irwin}, \& {Battaglia}}]{deboer2012_sculptor} {de Boer}, T.~J.~L., {Tolstoy}, E., {Hill}, V., et al. 2012{\natexlab{a}}, \aap, 539, A103
\bibitem[{{de Boer} {et~al.}(2012{\natexlab{b}}){de Boer}, {Tolstoy}, {Hill}, {Saha}, {Olszewski}, {Mateo}, {Starkenburg}, {Battaglia}, \& {Walker}}]{deboer2012} {de Boer}, T.~J.~L., {Tolstoy}, E., {Hill}, V., et al. 2012{\natexlab{b}}, \aap, 544, A73
\bibitem[{{Duquennoy} \& {Mayor}(1991)}]{duquennoy1991} {Duquennoy}, A. \& {Mayor}, M. 1991, \aap, 248, 485
\bibitem[{{Feroz} {et~al.}(2009){Feroz}, {Hobson}, \& {Bridges}}]{feroz2009} {Feroz}, F., {Hobson}, M.~P., \& {Bridges}, M. 2009, \mnras, 398, 1601
\bibitem[{{Fischer} \& {Marcy}(1992)}]{fischer1992} {Fischer}, D.~A. \& {Marcy}, G.~W. 1992, \apj, 396, 178
\bibitem[{{Fisher}(2004)}]{fisher2004} {Fisher}, R.~T. 2004, \apj, 600, 769
\bibitem[{{Goldberg} {et~al.}(2003){Goldberg}, {Mazeh}, \& {Latham}}]{goldberg2003} {Goldberg}, D., {Mazeh}, T., \& {Latham}, D.~W. 2003, \apj, 591, 397
\bibitem[{{Hurley-Keller} {et~al.}(1998){Hurley-Keller}, {Mateo}, \& {Nemec}}]{hurley-keller1998} {Hurley-Keller}, D., {Mateo}, M., \& {Nemec}, J. 1998, \aj, 115, 1840
\bibitem[{{King} {et~al.}(2012){King}, {Goodwin}, {Parker}, \& {Patience}}]{king2012} {King}, R.~R., {Goodwin}, S.~P., {Parker}, R.~J., \& {Patience}, J. 2012, \mnras, 427, 2636
\bibitem[{{Kohler} \& {Leinert}(1998)}]{kohler1998} {Kohler}, R. \& {Leinert}, C. 1998, \aap, 331, 977
\bibitem[{{Kroupa}(1995)}]{kroupa1995} {Kroupa}, P. 1995, \mnras, 277, 1491
\bibitem[{{Lee} {et~al.}(2009){Lee}, {Yuk}, {Park}, {Harris}, \& {Zaritsky}}]{lee2009} {Lee}, M.~G., {Yuk}, I.-S., {Park}, H.~S., {Harris}, J., \& {Zaritsky}, D.  2009, \apj, 703, 692
\bibitem[{{Leinert} {et~al.}(1993){Leinert}, {Zinnecker}, {Weitzel}, {Christou}, {Ridgway}, {Jameson}, {Haas}, \& {Lenzen}}]{leinert1993} {Leinert}, C., {Zinnecker}, H., {Weitzel}, et al. 1993, \aap, 278, 129
\bibitem[{{Lemasle} {et~al.}(2012){Lemasle}, {Hill}, {Tolstoy}, {Venn}, {Shetrone}, {Irwin}, {de Boer}, {Starkenburg}, \& {Salvadori}}]{lemasle2012} {Lemasle}, B., {Hill}, V., {Tolstoy}, E., et al. 2012, \aap, 538, A100
\bibitem[{{{\L}okas}(2009)}]{lokas2009} {{\L}okas}, E.~L. 2009, \mnras, 394, L102
\bibitem[{{{\L}okas} {et~al.}(2005){{\L}okas}, {Mamon}, \& {Prada}}]{lokas2005} {{\L}okas}, E.~L., {Mamon}, G.~A., \& {Prada}, F. 2005, \mnras, 363, 918
\bibitem[{{Marks} \& {Kroupa}(2012)}]{marks2012} {Marks}, M. \& {Kroupa}, P. 2012, \aap, 543, A8
\bibitem[{{Martinez} {et~al.}(2011){Martinez}, {Minor}, {Bullock}, {Kaplinghat}, {Simon}, \& {Geha}}]{martinez2011} {Martinez}, G.~D., {Minor}, Q.~E., {Bullock}, J., et al. 2011, \apj, 738, 55
\bibitem[{{Mayor} {et~al.}(1992){Mayor}, {Duquennoy}, {Halbwachs}, \& {Mermilliod}}]{mayor1992} {Mayor}, M., {Duquennoy}, A., {Halbwachs}, J.-L., \& {Mermilliod}, J.-C. 1992, in Astronomical Society of the Pacific Conference Series, Vol.~32, IAU Colloq. 135: Complementary Approaches to Double and Multiple Star Research, ed. H.~A. {McAlister} \& W.~I. {Hartkopf}, 73
\bibitem[{{Mazeh} {et~al.}(1992){Mazeh}, {Goldberg}, {Duquennoy}, \& {Mayor}}]{mazeh1992} {Mazeh}, T., {Goldberg}, D., {Duquennoy}, A., \& {Mayor}, M. 1992, \apj, 401, 265
\bibitem[{{Milone} {et~al.}(2012){Milone}, {Piotto}, {Bedin}, {Aparicio}, {Anderson}, {Sarajedini}, {Marino}, {Moretti}, {Davies}, {Chaboyer}, {Dotter}, {Hempel}, {Mar{\'{\i}}n-Franch}, {Majewski}, {Paust}, {Reid}, {Rosenberg}, \& {Siegel}}]{milone2012} {Milone}, A.~P., {Piotto}, G., {Bedin}, L.~R., et al. 2012, \aap, 540, A16
\bibitem[{{Minor} {et~al.}(2010){Minor}, {Martinez}, {Bullock}, {Kaplinghat}, \& {Trainor}}]{minor2010} {Minor}, Q.~E., {Martinez}, G., {Bullock}, J., {Kaplinghat}, M., \& {Trainor}, R. 2010, \apj, 721, 1142
\bibitem[{{Offner} {et~al.}(2009){Offner}, {Klein}, {McKee}, \& {Krumholz}}]{offner2009} {Offner}, S.~S.~R., {Klein}, R.~I., {McKee}, C.~F., \& {Krumholz}, M.~R. 2009, \apj, 703, 131
\bibitem[{{Olszewski} {et~al.}(1996){Olszewski}, {Pryor}, \& {Armandroff}}]{olszewski1996} {Olszewski}, E.~W., {Pryor}, C., \& {Armandroff}, T.~E. 1996, \aj, 111, 750
\bibitem[{{Paczy{\'n}ski}(1971)}]{paczynski1971} {Paczy{\'n}ski}, B. 1971, \araa, 9, 183
\bibitem[{{Patience} {et~al.}(2002){Patience}, {Ghez}, {Reid}, \& {Matthews}}]{patience2002} {Patience}, J., {Ghez}, A.~M., {Reid}, I.~N., \& {Matthews}, K. 2002, \aj, 123, 1570
\bibitem[{{Price} \& {Bate}(2010)}]{price2010} {Price}, D.~J. \& {Bate}, M.~R. 2010, in American Institute of Physics Conference Series, Vol. 1242, American Institute of Physics Conference Series, ed. G.~{Bertin}, F.~{de Luca}, G.~{Lodato}, R.~{Pozzoli}, \& M.~{Rom{\'e}}, 205--218
\bibitem[{{Raghavan} {et~al.}(2010){Raghavan}, {McAlister}, {Henry}, {Latham}, {Marcy}, {Mason}, {Gies}, {White}, \& {ten Brummelaar}}]{raghavan2010} {Raghavan}, D., {McAlister}, H.~A., {Henry}, T.~J., et al. 2010, \apjs, 190, 1
\bibitem[{{Richardson} \& {Fairbairn}(2013)}]{richardson2013} {Richardson}, T. \& {Fairbairn}, M. 2013, \mnras, 432, 3361
\bibitem[{{Scally} {et~al.}(1999){Scally}, {Clarke}, \& {McCaughrean}}]{scally1999} {Scally}, A., {Clarke}, C., \& {McCaughrean}, M.~J. 1999, \mnras, 306, 253
\bibitem[{{Skilling}(2004)}]{skilling2004} {Skilling}, J. 2004, in American Institute of Physics Conference Series, Vol.  735, American Institute of Physics Conference Series, ed. R.~{Fischer}, R.~{Preuss}, \& U.~V. {Toussaint}, 395--405
\bibitem[{{Smecker-Hane} {et~al.}(1996){Smecker-Hane}, {Stetson}, {Hesser}, \& {Vandenberg}}]{smecker-hane1996} {Smecker-Hane}, T.~A., {Stetson}, P.~B., {Hesser}, J.~E., \& {Vandenberg}, D.~A. 1996, in Astronomical Society of the Pacific Conference Series, Vol.~98, From Stars to Galaxies: the Impact of Stellar Physics on Galaxy Evolution, ed. C.~{Leitherer}, U.~{Fritze-von-Alvensleben}, \& J.~{Huchra}, 328--+
\bibitem[{{Sollima} {et~al.}(2007){Sollima}, {Beccari}, {Ferraro}, {Fusi Pecci}, \& {Sarajedini}}]{sollima2007} {Sollima}, A., {Beccari}, G., {Ferraro}, F.~R., {Fusi Pecci}, F., \& {Sarajedini}, A. 2007, \mnras, 380, 781
\bibitem[{{Sollima} {et~al.}(2012){Sollima}, {Bellazzini}, \& {Lee}}]{sollima2012} {Sollima}, A., {Bellazzini}, M., \& {Lee}, J.-W. 2012, \apj, 755, 156
\bibitem[{{Sterzik} {et~al.}(2003){Sterzik}, {Durisen}, \& {Zinnecker}}]{sterzik2003} {Sterzik}, M.~F., {Durisen}, R.~H., \& {Zinnecker}, H. 2003, \aap, 411, 91
\bibitem[{{Tonry} \& {Davis}(1979)}]{tonry1979} {Tonry}, J. \& {Davis}, M. 1979, \aj, 84, 1511
\bibitem[{{Walker} {et~al.}(2009{\natexlab{a}}){Walker}, {Mateo}, \& {Olszewski}}]{walker2009} {Walker}, M.~G., {Mateo}, M., \& {Olszewski}, E.~W. 2009{\natexlab{a}}, \aj, 137, 3100
\bibitem[{{Walker} {et~al.}(2007){Walker}, {Mateo}, {Olszewski}, {Bernstein}, {Sen}, \& {Woodroofe}}]{walker2007} {Walker}, M.~G., {Mateo}, M., {Olszewski}, et al. 2007, \apjs, 171, 389
\bibitem[{{Walker} {et~al.}(2009{\natexlab{b}}){Walker}, {Mateo}, {Olszewski}, {Sen}, \& {Woodroofe}}]{walker11-09} {Walker}, M.~G., {Mateo}, M., {Olszewski}, E.~W., {Sen}, B., \& {Woodroofe}, M.  2009{\natexlab{b}}, \aj, 137, 3109
\bibitem[{{Wolf} {et~al.}(2010){Wolf}, {Martinez}, {Bullock}, {Kaplinghat}, {Geha}, {Mu{\~n}oz}, {Simon}, \& {Avedo}}]{wolf2010} {Wolf}, J., {Martinez}, G.~D., {Bullock}, J.~S., et al. 2010, \mnras, 406, 1220
\bibitem[{{Yan} \& {Cohen}(1996)}]{yan1996} {Yan}, L. \& {Cohen}, J.~G. 1996, \aj, 112, 1489

\end{thebibliography}

\end{document}